\documentclass[journal]{IEEEtran}
\usepackage{cite}

\ifCLASSINFOpdf
  \usepackage[pdftex]{graphicx}
  \graphicspath{ {./figure/} }
  \DeclareGraphicsExtensions{.pdf,.jpg,.png,.eps}
\else
  \usepackage[dvips]{graphicx}
  \graphicspath{ {./figure/} }
  \DeclareGraphicsExtensions{.pdf,.jpeg,.png,.eps}
\fi
\usepackage{xcolor}
\usepackage{soul}

\usepackage{amsmath}
\usepackage{array}

\usepackage{multirow}

\usepackage{tikz}
\usepackage{pgfplots}
\usepackage{verbatim}

\usepackage{enumitem}   
\usepackage{fancyhdr}

\begin{document}
%
\title{Cancer Gene Profiling through Unsupervised Discovery}

%
%
%

\author{Enzo~Battistella,
        Maria~Vakalopoulou,~\IEEEmembership{Member,}
        Roger~Sun,
        Th\'eo~Estienne,
        Marvin~Lerousseau,
        Sergey Nikolaev,
        \'Emilie~Alvarez~Andres,
        Alexandre~Carr\'e,
        St\'ephane~Niyoteka,
        Charlotte~Robert,
        Nikos~Paragios,~\IEEEmembership{Fellow Member},
        and~\'Eric~Deutsch
}
\maketitle


\begin{abstract}
Precision medicine is a paradigm shift in healthcare relying heavily on genomics data. However, the complexity of biological interactions, the large number of genes as well as the lack of comparisons on the analysis of data, remain a tremendous bottleneck regarding clinical adoption. 
In this paper, we introduce a novel, automatic and unsupervised framework to discover low-dimensional gene biomarkers. 
Our method is based on the LP-Stability algorithm, a high dimensional center-based unsupervised clustering algorithm, that offers modularity as concerns metric functions and scalability, while being able to automatically determine the best number of clusters. Our evaluation includes both mathematical and biological criteria. The recovered signature is applied to a variety of biological tasks, including screening of biological pathways and functions, and characterization relevance on tumor types and subtypes. Quantitative comparisons among different distance metrics, commonly used clustering methods and a  referential gene signature used in the literature, confirm state of the art performance of our approach. In particular, our signature, that is based on $27$ genes, reports at least $30$ times better mathematical significance (average Dunn's Index) and $25\%$ better biological significance (average Enrichment in Protein-Protein Interaction) than those produced by other referential clustering methods. Finally, our signature reports promising results on distinguishing immune inflammatory and immune desert tumors, while  reporting a high balanced accuracy of $92\%$ on tumor types classification and averaged balanced accuracy of $68\%$ on tumor subtypes classification, which represents, respectively $7\%$ and $9\%$ higher performance compared to the referential signature.

\end{abstract}


%
\IEEEpeerreviewmaketitle

\section{Introduction}
\footnotetext{Version submitted to IEEE Transactions  on Bioinformatics and Computational Biology© 2021 IEEE.  Personal use of this material is permitted.  Permission from IEEE must be obtained for all other uses, in any current or future media, including reprinting/republishing this material for advertising or promotional purposes, creating new collective works, for resale or redistribution to servers or lists, or reuse of any copyrighted component of this work in other works.}
%
%
%
%
\IEEEPARstart{O}{mics}  data analysis - including genomics, transcriptomics and metabolomics - has greatly benefited from the tremendous sequencing technique advances~\cite{Kurian2014} allowing to highly increase the quality and the quantity of data. These omics techniques are pivotal aspects of the development of personalized medicine by enabling a better understanding of fine-grained molecular mechanisms~\cite{Hanahan2011}. In oncology, these techniques provide a more comprehensive insight of the biological processes intricacy in cancers giving momentum to molecular-type characterization through omics or even multi-omics approaches~\cite{ramaswamy2001multiclass,Chen2017}. Such a precise and robust characterization is a highly valuable asset for tumor characterization and provides significant acumen on their treatment.

Genomics, probably the most prominent omics technique, refer to the study of entire genomes contrary to genetics that interrogate individual variants or single genes~\cite{Hasin2017}. In this direction, novel methods study specific variants of genes aimed at producing robust biomarkers, which contribute to both the response of patients to treatment~\cite{wan2010,Sun2018} and the association with complex and Mendelian diseases~\cite{Dunne2017}. However, the relatively low number of samples per tumor subtype, along with the curse of dimensionality and the lack of ground truth affect many of these studies~\cite{Drucker2013}, which may prevent any statistically meaningful causal relation discovery. 

Unsupervised clustering is a very efficient technique to study large high-dimensional datasets aimed at discovering unknown indiscernible structures and correlations~\cite{Halkidi2001}. Clustering algorithms aspire to single out a group separation of the data favoring low variation inside the groups and high variation between groups. Notwithstanding, there is a large variety of clustering approaches, relying on different properties, leading to significantly different outcomes. 
The main challenge of clustering is the definition of a metric/similarity function depicting the notion of closeness between objects under consideration. This includes not only the intrinsic clustering properties the algorithm seeks to optimize but also the notion of distance involved. Qualitative and quantitative evaluation is a critical step towards clustering effective adoption and relies on the basis of independent and reliable measures for the proper comparison of the parameters and methods. Numerous existing  metrics  assess the quality of the clusters from a statistical point of view as the Silhouette Value~\cite{Kaufman1987}, Dunn's Index~\cite{Kovacs2005} or more recently the Diversity Method~\cite{Kingrani2017}. In addition, in presence of annotations, the Rand Index~\cite{hubert1985}. Clustering evaluation is even more challenging in the case of genomics, as the clusters should also be biologically informative. Protein-Protein Interaction (PPI) and the Gene Ontology (GO) have been recently introduced in this sub-domain to assess the biological soundness of the clusters through Enrichment Scores~\cite{Wagner2015,Pepke2017}.

\begin{figure*}[!t]
\centering
\includegraphics[scale=0.4]{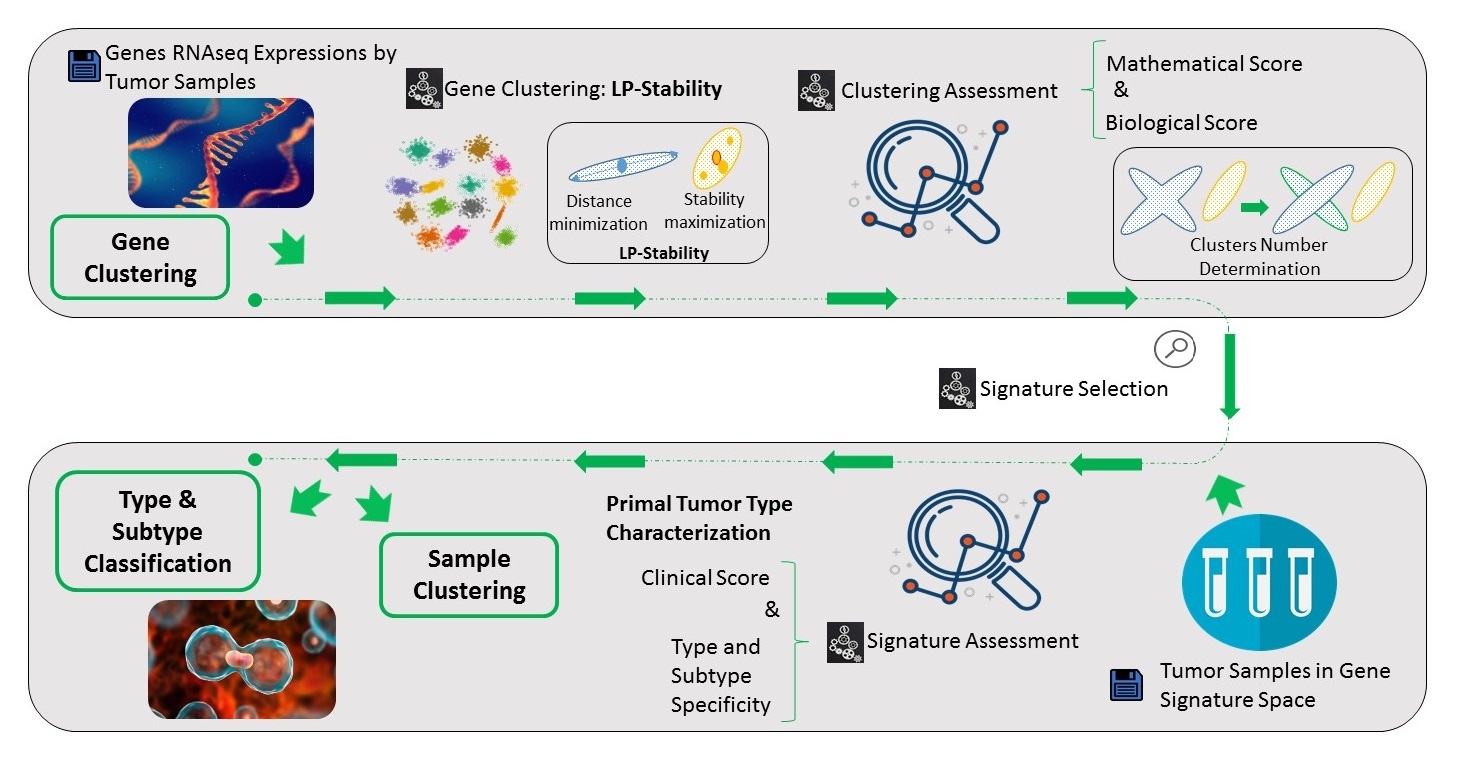}
\caption{\textbf{Proposed Framework.}
      A general overview of the different steps of our process. Our proposed framework is composed of two steps. First, a clustering algorithm, here LP-Stability, is used to generate clusters of genes having similar expression profiles. Then, the clustering that performs best on both mathematical and biological scores is selected as a gene signature. In the second step, the generated signature is used to perform sample clustering and sample classification. The performance on this step is evaluated by analysing the distribution of the samples into the different clusters or the performance on the classification tasks, here the target was the tumor types and subtypes characterization.}
\label{fig:pipeline}
\end{figure*}

The use of cluster analysis on RNA-seq transcriptomes is a wide-spread technique~\cite{Cowen2017} whose main goal is to define groups of genes that have similar expression profiles, proposing compact signatures~\cite{Dunne2017}. These robust signatures are necessary to identify associations with different biological processes, as tumor types or cancer molecular subtypes, and to highlight gene coding for proteins interacting together or participating in the same biological process~\cite{vanDam2017}. 
 The main advantages of unsupervised clustering compared to supervised approaches - or methods guided by specific biological functions or processes - is the ability to discover unknown patterns, associations and correlations in the genome. Furthermore, unsupervised discovery offers better tractability when applied to the tremendous amount of genes. In addition, studies that perform clustering relying on a priori knowledge lead to redundant signatures and loss of information~\cite{Cantini2017}. This is one of the reasons that several studies focus on statistical pattern recognition
methods such as the center-based K-Means~\cite{Macqueen1967}, the model-based CorEx~\cite{VerSteeg2014} or the stability-based LP-Stability~\cite{Komodakis2009} towards the identification of meaningful and predictive groups of genes as biomarkers~\cite{Bailey2018}.  
In this direction, CorEx~\cite{Pepke2017} has recently been introduced to generate gene signatures evaluated and optimized over ovarian tumors that address this specific tumor type with a high dimensional gene signature composed of several hundred genes. To deal with this high dimensionality, studies propose methods to combine and prune existing signatures to obtain a unique compact and informative signature~\cite{Cantini2017,thorsson2018}.

Although dimension reduction through clustering is not new~\cite{Pepke2017}, there is an important shortfall in literature of a thorough, mathematically and biologically meaningful comparison of clusterings methods on a same database. In many studies, a single evaluation metric is used and there is no relevant comparison with other algorithms. By “relevant”, we mean here that the optimization of the different baseline algorithm hyperparameters is ensured and compared through a fair evaluation metric. Mathematical metrics for instance, are highly dependent on the property the algorithm is optimizing and the distance notion considered. The evaluation of this bias through, as an example, random clusters using different distance notions to offer a fair comparison between the different algorithms. Finally, a few surveys~\cite{Oyelade2016} propose a thorough comparison, using several evaluation criteria, albeit reporting results shown in several other studies without actually comparing the methods on a same database with all the criteria at once.

In this paper,  we introduce a novel unsupervised approach that is  modular, scalable and metric free towards the definition of a predictive gene signature while proposing a complete methodology for comparison, analysis and evaluation of genomic signatures. The backbone of our methodology refers to  a powerful graph-based unsupervised clustering method, the LP-Stability algorithm~\cite{Komodakis2009}, which has been successfully adapted in various fields including, recently, in genomics~\cite{battistella2019}.
Our approach offers: 

\begin{enumerate}[label=(\roman*)]
    \item Standardization and automatization concerning gene clustering evaluation for the selection of the best distance notions, metrics, algorithms and hyperparameters; 
    \item Creation of generic, low dimensional signatures using the gene expressions of all coding genes, including comparisons to random signatures to highlight statistical superiority;
    \item Systematic assessment of the biological power of gene signatures by evaluating the different tumor type and subtype associations via supervised (proving tissue-specificity and predictive power), and unsupervised (proving automatic discovery and expression power) techniques. By this, we demonstrated the power of the proposed gene signature (based on $27$ genes) compared to other methods in the literature;
    \item Thorough biological analysis of the processes involved in sample clusters via gene screening techniques, affirming the robustness of the obtained results. 

\end{enumerate}

\section{Overview of the Proposed Approach}

The overview of the method presented in this paper is summarized in Fig.~\ref{fig:pipeline}. 
To evaluate and select the best gene signature, we introduce two distinct metrics checking both mathematical and biological properties. In particular, we used the mathematical assessment metric of Dunn's Index (DI)~\cite{Kovacs2005} and the biological one of Enrichment Score in PPI~\cite{Pepke2017} which are both referential for the assessment of clustering although they have never been combined.
Then, a low dimensional aggregated gene signature is defined by combining representative genes in each cluster. To prove the power of the discovered biomarker, a systematic and thorough evaluation regarding its biological and clinical relevance was performed. In particular, the signature was evaluated and compared through sample clustering and sample classification. As targeted by the sample clustering, we chose the different tumor types and assessed the success of the clustering through sample distribution analysis and clustering evaluation metrics such as Rand Index and Mutual Information. In addition, we used the method from~\cite{tusher2001significance} to obtain important genes for the samples of each cluster which were associated to their pathways using~\cite{szklarczyk2018}. Finally, the last evaluation criteria was the performance in categorizing the cancer types and subtypes through supervised machine learning techniques. Our proposed signature has been compared against both signatures designed from commonly used algorithms for gene clustering~\cite{Macqueen1967,Pepke2017} and a recently proposed prominent gene signature~\cite{thorsson2018}.

\section{Discovering Correlations in Gene Expressions}
\label{section:methodo}

\subsection{Algorithms}
We consider here $n$ points $S=\{x^1,...x^n\}$ in a space of dimension $m$ where each point $x^p$ coordinates will be described by $x^p=(x^p_1,...,x^p_m)$.
 In this study, we consider several notions of dissimilarity $d$.
If we denote $k$ the number of clusters in a clustering $C$, then the clustering is a set of clusters $C=\{C_1,...,C_k\}$ defined such that $\forall 1\leq i,j \leq k$, $C_i \cap C_j = \emptyset$ and $\bigcup_{1\leq i\leq k}  C_i = S$. 
The number of points in cluster $C_i$ will be denoted by $n_i$.
We call centroid $\mu_i$ of cluster $C_i$ the mean of the points of the cluster.
Finally, we get a discrete random variable $X=\{X_1,...,X_b\}$ from a point $x\in S$ by binning $b$ bins. We denote $P(X)$ the probability mass function of $X$. Then from that, the Shannon Entropy $H$ of variable $X$ is defined by $H(X)=-\sum_{1\leq i \leq b} P(X_i)\ln P(X_i)$.
 


\textbf{K-Means algorithm}~\cite{Macqueen1967} is a very popular and simple algorithm used for data following Gaussian distributions. First, the algorithm draws an initial random set of cluster centroids. Then, until convergence it iteratively determines $k$ clusters by assigning the points to their closest centroid and computes their new centroids $\mu_i$. The algorithm aims to solve

\begin{equation}
 \min_C \sum_{i=1}^k \sum_{x\in C_i} d(x, \mu_i) \; .
\end{equation} 
Only the number of clusters $k$ has to be defined beforehand. 
Generally, K-Means is used with Euclidean distance for convergence issues. A main drawback of this technique is that the random initialization is a source of nondeterminism and may cause instability in the cluster generation for different runs. We address this issue by selecting the best clustering over multiple iterations with a different initialization.

\textbf{CorEx algorithm}~\cite{VerSteeg2014} is a model-based algorithm that has been applied to various fields and, especially on genes~\cite{Pepke2017} with great success. This algorithm aims to define a set $S'$ of $k$ latent factors accounting for the most variance of the dataset $S$. Formally, it relies on the Total Correlation of discrete random variables $X^1,...,X^p$ defined by 

\begin{equation}
TC(X^1,...,X^p)=\sum_{1 \leq i \leq p} H(X^i) - H(X^1,...,X^p)
\end{equation}

and the Mutual Information of two random variables 

\begin{equation}
MI(X^i,X^j)= \sum_{X^i_p\in X^i}\sum_{X^j_q\in X^j}P(X^i_p,X_i^q)\log\dfrac{P(X^i_p,X^j_q)}{P(X^i_p)P(X^j_q)}
\end{equation}
where $P(X^i_p,X^j_q)$ is the joint probability function and $P(X^i_p),P(X^j_q)$ are  marginal probability functions. To guarantee a reliable definition of the latent variables, the algorithm minimizes the Total Correlation, $TC(S|S')$, corresponding to the additional information brought by the points in $S$ compared to the latent factors of $S'$. Then, to obtain the clustering, each point $x^p$ is allocated to the cluster of the latent factor $f$ maximizing the mutual information, $MI(X^p,f)$. 
Similar to K-Means, the only requirement of CorEx algorithm is the number of clusters $k$. Moreover, K-Means is the only algorithm under consideration that directly operates on the gene expression matrix to build a statistical model without using a given distance matrix. 


\textbf{LP-Stability algorithm}~\cite{Komodakis2009} is based on linear programming and relies on the same definition of clusters as K-Means \textit{i.e.} we want to minimize the distance between each point of a cluster and the center of the cluster. However, the novelty and interest of this technique is that instead of taking centroids as cluster centers, it defines stable cluster centers. 
Formally, we aim to optimize the following linear system 

\begin{align}
\begin{split}
 PRIMAL ~\equiv ~& \min_C \sum_{p,q} d(x^p,x^q)C(p,q) \\
			 & s.t.~ \sum_q C(p,q)=1 \\
			 & ~ ~~~C(p,q) \leq C(q,q) \\
			 & ~ ~~~C(p,q) \geq 0 \; .
\end{split}
\end{align}
where $C(p,q)$ represents the fact that $x^p$ belongs to the cluster of center $x^q$. The formula corresponds to the minimization of the distance between a point and its cluster center while ensuring that each  point must belong to one and only one cluster and that centers belong to their own cluster. The determination of the stable centers relies on the following notion of stability:
\begin{align*}
S(q)=\inf_{s} \{ s, \ d(q,q) + s ~ \ \text{PRIMAL has no }\\ \text{optimal solution with} \ C(q,q)>0 \} \; .
\end{align*}
The stability of a point is the maximum penalty the point can receive while remaining an optimal cluster center in PRIMAL. Besides, to better exploit particular field constraints of the points or to better tune the number of clusters, penalty value $S_q>=0$ can be added to point $q$. We will then consider the penalty vector $S$ weighting the distance $d$ such as $\forall q, S_q\in S, d'(q,q)=d(q,q) + S_q$. Doing so, we will impose a stronger minimal stability for the cluster centers entailing a lower number of clusters.


Let us denote $\mathcal{Q}$ the set of stable clusters centers. The algorithm solves the clustering using the DUAL problem 

\begin{align}
\begin{split}
 DUAL ~\equiv ~& \max_D D(h)=\sum_{p\in \mathcal{V}} h^p \\
			 & s.t.~ h^p = \min_{q\in \mathcal{V}}h(p,q) \\
			 & ~ ~~~\sum_{p\in \mathcal{V}} h(p,q)=\sum_{p\in \mathcal{V}} d(x^p,x^q) \\
			 & ~ ~~~h(p,q) \geq d(x^p,x^q) \; .
\end{split}
\end{align}
where $h(p,q)$ corresponds here to the minimal pseudo-distance between $x^p$ and $x^q$ and $h^p$ to the one from $x^p$.
This previous DUAL problem is then conditioned by considering only centers in the set of stable points $\mathcal{Q}$:

\begin{equation}
DUAL_\mathcal{Q} = \max DUAL ~ s.t. ~h_{pq}=d_{pq}, \forall\{p,q\}\cap \mathcal{Q} \neq \emptyset \; .
\end{equation}

This method presents several advantages. It is versatile and can integrate any metric function while, it does not make prior assumptions on the number of clusters or their distribution. It aims to define clustering in a global manner seeking for an automatic selection of the cluster centers.  For that matter, it relies on the optimization of the set of stable centers, as well as the assignment of each observation to the most appropriate cluster, meaning the one minimizing the distance to the center.  
This algorithm only requires a penalty vector $S$, influencing the number of clusters.

\subsection{Proximity Measures}
\label{section:dist}
To tackle the issue of high dimensionality of the data combined with a low ratio between samples and dimensions of each sample, we studied several different distance notions.
In this study we considered the following distances: the Euclidean distance, the Cosine distance, the Pearson's correlation, the Spearman's rank correlation, the Kendall's rank correlation and the Kullback-Leibler divergence. These standard metrics are further detailed in Supplementary Materials.

Depending on the type of data or algorithm used, different proximity measures may heavily impact the performance for the clustering.
The different correlations cover the range of $[-1,1]$, the value is positive when the observations evolve in a similar way for the compared variables and negative when they evolve in opposite ways. 
High absolute values indicate high correlations in the observations. On the other hand, high values in terms of distance indicate observations that are not similar in the specific feature space.
In this work, we used 
$\sqrt{2(1 - c)}$ 
to convert correlations into distances. 
  For simplicity, distances coming from correlations will be referred to as correlation-based distances for the rest of the paper.

\subsection{Unsupervised Gene Clustering Evaluation}
To evaluate the performance of gene clustering methods, we used both mathematical and biological criteria. The quality of the results was assessed using the biological relevance information brought by the Enrichment Score, while the prominent Dunn's Index statistical method was considered regarding the clustering mathematical appropriateness. 

\begin{itemize}
    \item \textbf{Enrichment Score (ES):} Enrichment is the most commonly adopted technique to assess biological relevance  in an automatic manner~\cite{Pepke2017}.  We considered here the Enrichment in PPI by studying the proteins corresponding to the genes.  Even if, contrary to enrichment in a given biological process, PPI  does not integrate specific information about predefined pathways and  biological processes, it fulfills our aim of an unbiased and general metric. Enrichment for a cluster represents the probability of obtaining the same number of interactions in a random set of genes of same size as in the evaluated cluster. 
    In particular, the cluster is considered as enriched if the p-value is below a given threshold (abbreviated by th).
    The ES corresponds to the proportion of enriched clusters in the clustering. To calculate the ES, Stringdb library based on String PPI network~\cite{szklarczyk2018} was used.
    
    \item \textbf{Dunn's Index (DI):} The Dunn's Index~\cite{Kovacs2005} studies the ratio between inter-cluster and intra-cluster variance. The former is meant to be large as the distributions in different clusters should be different. The latter has to be small as we want points that are in a same cluster to follow a common distribution. Formally,\\ $Dunn(\mathcal{C}) = \dfrac{\min_{1\leq i,j \leq k} \delta(C_i,C_j)}{\max_{1 \leq i \leq k}\Delta(C_i)}$ where $\delta(C_1,C_2)$ is the distance between the two closest points of the clusters $C_i$ and $C_j$, $\Delta(C_i)$ is the diameter of the cluster  \textit{i.e.} the distance between the two farthest points of the cluster $C_i$.
    This assessment score is highly sensitive to extreme not well-formed clusters making it ideal for our problem.
    
\end{itemize}

\section{Definition of a Low Dimensional Cancer Signature}
\label{section:methodo2}


\subsection{Signature Selection}
\label{section:ss}
The ultimate goal of our approach is to produce low dimensional signatures as a byproduct of unsupervised clustering outcome. To this end, the signatures were produced by selecting the most representative gene per cluster for the clusterings with the highest ES and DI performance. For LP-Stability algorithm, the selected genes were the stable centers that the algorithm relies on, in the the rest of the algorithms, we chose as representatives the clusters medoids which are the genes the closest to the cluster centroid. This choice is motivated by the fact that the stable centers of the clustering obtained through LP-Stability are also medoids.

In complement, once a signature is selected, a redundancy analysis was performed using STRING tool~\cite{szklarczyk2018} to decipher any biological process that was particularly over-represented so suggesting redundancy of the information.
 In addition, Genotype-Tissue Expression (GTEx) portal~\footnote{www.gtexportal.org} was used to assess the tissue specificity for the proposed signature. A good signature should present genes with different expression profiles over the different tissues. This tool offers a visual representation for each gene of their expression and regulation in different tissues. It relies on the analysis of multiple human tissues from donors to identify correlations between genotype and tissue-specific gene expression levels.


\subsection{Sample Clustering: Discovery Power}
\label{section:DP}
In order to perform sample clustering, we compared several algorithms and distances. The most meaningful results were obtained with the K-Medoids method, a variant of K-Means, combined with the Spearman's rank correlation-based distance.  The relevance of the obtained results was assessed by analyzing the partition of the different tumor types in the clusters.
In particular, driven from known biological evidence, we considered as meaningful the associations of lung tumors (LUSC, LUAD), squamous tumors (LUSC, HNSC, CESC), gynecologic tumors (BRCA, OV, CESC), smoking related tumors (LUSC, LUAD, BLCA, CESC, HNSC). We disregarded samples types in a cluster representing less than $5\%$ of the total cluster size. We considered a poorly defined cluster to be a cluster  presenting less than $50$ samples or distribution of the samples types in the same proportions as in the whole dataset as it would show random associations.

Gene screening analysis was also used to identify the genes that are expressed differently over the sample clusters and thus indicating the biological processes involved. For that, we used the SAM method~\cite{tusher2001significance}, that aims to identify the genes that are differently expressed over two groups of samples. SAM assesses the significance of the variations of the gene expression using a statistical t-test, providing a significance score and a False Discovery Rate (FDR). 
To better assess the relevance of separating samples of the same tumor type, we studied the genes that are expressed differently for each tumor type in a cluster compared to all the other samples of the same tumor type.
We thus pinpointed significant genes for each cluster and each tumor type by cluster. Once more, the method in~\cite{szklarczyk2018} has been used for assessing the biological relevance of the clusters and their association to different tumor types, by studying the biological processes involved. A well-defined sample clustering is characterized by different clusters presenting different enriched biological processes and pathways while different tumor types in a same cluster should be enriched in the same ones.

\subsection{Sample Clustering: Expression Power}
To assess how well the different tumor types have been separated, we used several different metrics (formal definitions are provided in Supplementary Materials).
\begin{itemize}
    \item \textbf{Adjusted Rand Index (ARI)} is a similarity measure between a clustering $C'$ and a ground truth $C$. ARI corresponds to the proportion of pairs of elements that are in different clusters in both $C$ and $C'$ called $a$ or in a same cluster in both $C$ and $C'$ called $b$. 
    
    \item \textbf{Normalized Mutual Information (NMI)}  is a normalized measure of the mutual dependence between a clustering and the reference group.
    \item \textbf{Homogeneity:} Considering a clustering $C'$ and a ground truth $C$. It values clusters of $C$ containing elements all belonging to a same cluster in $C'$.
    \item \textbf{Completeness:} Considering a clustering $C'$ and a ground truth $C$. This complement of homogeneity values clusters of $C$ having all their elements belonging to a same cluster in $C'$.
    
    \item \textbf{Fowlkes-Mallow Score (FMS):}  It corresponds to the geometric mean of the pairwise precision and recall.

\end{itemize}

\subsection{Supervised Tumor Types/SubTypes Categorization}
The evaluation of the provided signatures were further assessed by a supervised setting in order to highlight their tissue specificity properties.
The supervised framework for tumor types and subtypes categorization was adapted from~\cite{covid1}. 
This classification pipeline relies on an ensemble of machine learning classifiers, exploring the ones with strong generalisation power. The best performing in terms of balanced accuracy and generalisation are combined through a probabilistic consensus schema to provide the appropriate label.

Towards the evaluation of the reported performance, we relied on classic machine learning metrics, namely balanced accuracy, weighted precision, weighted specificity and weighted sensitivity. The use of weighted metrics instead of the non-weighted ones is required here as we are considering a multi-class classification task with very unbalanced classes. The weighted scores (WS) were defined as

\begin{equation}
    WS = \frac{1}{N}\sum_l N_l S_l
\end{equation}

where $N$ corresponds to the total number of samples, $N_l$ the number of samples with class of label $l$ and $S_l$ the non-weighted score in one-vs-rest classification for the class $l$.

\section{Implementation Details}
\label{section:ID}
The parameters of each algorithm  for the gene clustering were obtained using grid search. In order to benchmark the behavior of each algorithm on different number of clusters, we evaluated their performance for the following number of clusters: from $2$ to $10$ with an increment of $1$, $15$, $20$, $25$ and between $30$ and $100$ with an increasing step of $10$ for the Random Clustering and K-Means algorithms and $25$ for CorEx algorithm because of its computational complexity. LP-Stability  automatically determines the number of clusters. In order to create meaningful comparisons, we adjusted the penalty vector $S$ in order to obtain approximately the same number of clusters as with the rest of the algorithms. For comparison purposes, we used the same penalty for all the genes, however, for the LP-Stability algorithm the penalty value could be adjusted and customized depending on the importance of specific genes.  

For the ES we reported the behavior of the algorithms with different threshold values \textit{i.e.} $0.005$, $0.025$, $0.05$ and $0.1$. Furthermore, in reporting the DI value, each method has been evaluated with the same proximity measure it relies on. For K-Means that is sensitive to initialization, we performed $100$ iterations for each parameter and selected the best clustering based on DI only to cope with the computational cost of the ES. This iterative process augments the computational time of the algorithm, but reports clusters with better statistical significance and more stable scores. Similarly, we performed $100$ repetitions of random clustering and observed rather similar results, we selected the clustering that reports the best DI score and reported its results.

For the sample clustering, we considered $10$ clusters corresponding to the actual $10$ tumor types. For the gene screening, we selected the most significant genes that reported a significance score of $7$ which corresponds to a q-value of FDR close to zero in most cases, while for the biological processes we considered only the $10$ most enriched processes by screening. 


\begin{table}[!t]
\caption{\textbf{Description of the dataset used in this study.} The different tumors and tumor types together with the corresponding number of samples are summarised. Urothelial  Bladder  Carcinoma  (BLCA),  Breast  Invasive Carcinoma (BRCA), Cervical Squamous Cell Carcinoma  and  Endocervical  Adenocarcinoma  (CESC),
Glioblastoma  Multiforme  (GBM),  Head  and  Neck Squamous  Cell  Carcinoma  (HNSC),  Liver  Hepatocellular Carcinoma (LIHC), Rectum Adenocarcinoma (READ), Lung adenocarcinoma (LUAD), Lung Squamous  Cell  Carcinoma  (LUSC)  and  Ovarian Cancer (OV).}
\label{tab:tumors-locations}
\centering
\begin{tabular}{|c|c|c|c|c|}
\hline
\multirow{2}{*}{Tumor Type} & Clustering            & \multicolumn{3}{c|}{Classification}                              \\ \cline{2-5} 
                            & \#Samples             & \#Samples Types       & \multicolumn{2}{c|}{\#Samples Subtypes:} \\ \hline
BLCA                        & 427                   & 129                   & \multicolumn{2}{c|}{——}                  \\ \hline
\multirow{5}{*}{BRCA}       & \multirow{5}{*}{1212} & \multirow{5}{*}{1223} & Normal:                   & 144          \\ \cline{4-5} 
                            &                       &                       & LumA:                     & 582          \\ \cline{4-5} 
                            &                       &                       & LumB:                     & 220          \\ \cline{4-5} 
                            &                       &                       & Her2:                     & 83           \\ \cline{4-5} 
                            &                       &                       & Basal:                    & 194          \\ \hline
CESC                        & 309                   & ——                    & \multicolumn{2}{c|}{——}                  \\ \hline
GBM                         & 171                   & 827                   & \multicolumn{2}{c|}{——}                  \\ \hline
\multirow{4}{*}{HNSC}       & \multirow{4}{*}{566}  & \multirow{4}{*}{279}  & Mesenchymal:              & 75           \\ \cline{4-5} 
                            &                       &                       & Basal:                    & 87           \\ \cline{4-5} 
                            &                       &                       & Atypical:                 & 68           \\ \cline{4-5} 
                            &                       &                       & Classical:                & 49           \\ \hline
\multirow{3}{*}{LIHC}       & \multirow{3}{*}{423}  & \multirow{3}{*}{183}  & iCluster1:                & 65           \\ \cline{4-5} 
                            &                       &                       & iCluster2:                & 55           \\ \cline{4-5} 
                            &                       &                       & iCluster3:                & 63           \\ \hline
LUAD                        & 576                   & 230                   & \multicolumn{2}{c|}{——}                  \\ \hline
LUSC                        & 552                   & 178                   & \multicolumn{2}{c|}{——}                  \\ \hline
\multirow{4}{*}{OV}         & \multirow{4}{*}{307}  & \multirow{4}{*}{489}  & Proliferative:            & 138          \\ \cline{4-5} 
                            &                       &                       & Mesenchymal:              & 109          \\ \cline{4-5} 
                            &                       &                       & Differentiated:           & 135          \\ \cline{4-5} 
                            &                       &                       & Immunoreactive:           & 107          \\ \hline
\multirow{2}{*}{READ}       & \multirow{2}{*}{72}   & \multirow{2}{*}{111}  & CIN:                      & 102          \\ \cline{4-5} 
                            &                       &                       & GS:                       & 9            \\ \hline
\end{tabular}
\end{table}

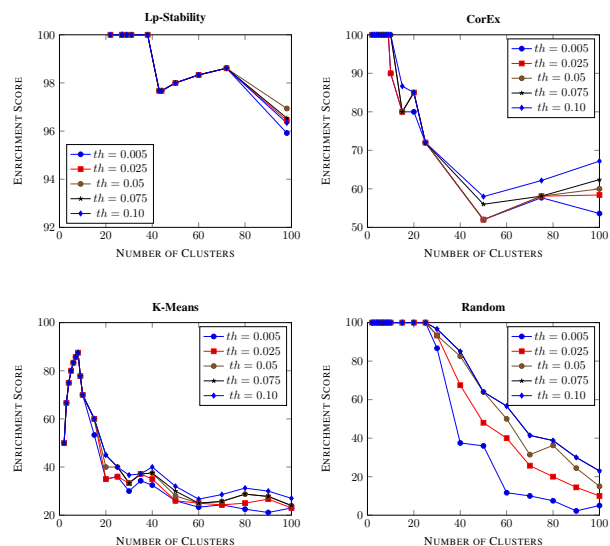
\begin{figure}[!b]
\centering
\scalebox{0.45}{
\begin{tikzpicture}
    \begin{axis}[
        title={\textbf{Lp-Stability}},
        xlabel=\textsc{Number of Clusters},
        ylabel= \textsc{Enrichment Score},
        xmin=0,xmax=100,ymin=92,ymax=100,
        legend style={at={(0.03,0.02)},anchor=south west}
    ]
            \addplot plot coordinates {
        (22,     100)
        (27,    100)
        (27,    100)
        (29,   100)
        (31,   100)
        (38,   100)
        (43,  97.6744186)
        (44,  97.6744186)
        (50,  98)
        (60,  98.33333333)
        (72,  98.61111111)
        (98,  95.91836735)
    };
        \addplot plot coordinates {
        (22,     100)
        (27,    100)
        (27,    100)
        (29,   100)
        (31,   100)
        (38,   100)
        (43,  97.6744186)
        (44,  97.6744186)
        (50,  98)
        (60,  98.33333333)
        (72,  98.61111111)
        (98,  96.43877551)
    };
      \addplot plot coordinates {
        (22,     100)
        (27,    100)
        (27,    100)
        (29,   100)
        (31,   100)
        (38,   100)
        (43,  97.6744186)
        (44,  97.6744186)
        (50,  98)
        (60,  98.33333333)
        (72,  98.61111111)
        (98,  96.93877551)
    };

    \addplot plot coordinates {
        (22,     100)
        (27,    100)
        (27,    100)
        (29,   100)
        (31,   100)
        (38,   100)
        (43,  97.6744186)
        (44,  97.6744186)
        (50,  98)
        (60,  98.33333333)
        (72,  98.61111111)
        (98,  96.53877551)
    };

    \addplot plot coordinates {
        (22,     100)
        (27,    100)
        (27,    100)
        (29,   100)
        (31,   100)
        (38,   100)
        (43,  97.6744186)
        (44,  97.6744186)
        (50,  98)
        (60,  98.33333333)
        (72,  98.61111111)
        (98,  96.33877551)
    };

    \legend{$th=0.005$\\$th= 0.025$\\$th=0.05$\\$th=0.075$\\$th=0.10$\\}
 \end{axis}
\end{tikzpicture}}
\scalebox{0.45}{
\begin{tikzpicture}
    \begin{axis}[
        title={\textbf{CorEx}},
        xlabel=\textsc{Number of Clusters},
        ylabel= \textsc{Enrichment Score},
        xmin=0,xmax=100,ymin=50,ymax=100,
        legend style={anchor=north east}
    ]
              \addplot plot coordinates {
        (2,     100)
        (3,    100)
        (4,    100)
        (5,   100)
        (6,   100)
        (7,   100)
        (8,  100)
        (9,  100)
        (10,  90)
        (15,  80)
        (20,  80)
        (25,  72)
        (50, 52)
        (75, 57.703)
        (100, 53.6)
    };
          \addplot plot coordinates {
        (2,     100)
        (3,    100)
        (4,    100)
        (5,   100)
        (6,   100)
        (7,   100)
        (8,  100)
        (9,  100)
        (10,  90)
        (15,  80)
        (20,  85)
        (25,  72)
        (50, 52)
        (75, 58.1081)
        (100, 58.4)
    };
      \addplot plot coordinates {
        (2,     100)
        (3,    100)
        (4,    100)
        (5,   100)
        (6,   100)
        (7,   100)
        (8,  100)
        (9,  100)
        (10,  100)
        (15,  80)
        (20,  85)
        (25,  72)
        (50, 52)
        (75, 58.1081)
        (100,60)
    };
 \addplot plot coordinates {
        (2,     100)
        (3,    100)
        (4,    100)
        (5,   100)
        (6,   100)
        (7,   100)
        (8,  100)
        (9,  100)
        (10,  100)
        (15,  80)
        (20,  85)
        (25,  72)
        (50, 56)
        (75, 58.1081)
        (100,62.4)
    };
      \addplot plot coordinates {
        (2,     100)
        (3,    100)
        (4,    100)
        (5,   100)
        (6,   100)
        (7,   100)
        (8,  100)
        (9,  100)
        (10,  100)
        (15,  86.66)
        (20,  85)
        (25,  72)
        (50, 57.99)
        (75, 62.16)
        (100, 67.2)
    };

    \legend{$th=0.005$\\$th= 0.025$\\$th=0.05$\\$th=0.075$\\$th=0.10$\\}
 \end{axis}
\end{tikzpicture}}

\vspace{5mm}
\scalebox{0.45}{
\begin{tikzpicture}
    \begin{axis}[
        title={\textbf{K-Means}},
        xlabel=\textsc{Number of Clusters},
        ylabel= \textsc{Enrichment Score},
        xmin=0,xmax=100,ymin=20,ymax=100,
        legend style={anchor=north east}
    ]

            \addplot plot coordinates {
        (2,     50)
        (3,    66.6667)
        (4,    75)
        (5,   80)
        (6,   83.333)
        (7,   85.71)
        (8,   87.5)
        (9,   77.7778)
        (10,  70)
        (15,  53.333)
        (20,  35)
        (25,  36)
        (30, 30)
        (35, 34.29)
        (40, 32.5)
        (50, 26)
        (60, 23.333)
        (70, 24.28)
        (80, 22.5)
        (90, 21.111)
        (100, 23)
    };
       \addplot plot coordinates {
        (2,     50)
        (3,    66.6667)
        (4,    75)
        (5,   80)
        (6,   83.333)
        (7,   85.71)
        (8,   87.5)
        (9,   77.7778)
        (10,  70)
        (15,  60)
        (20,  35)
        (25,  36)
        (30, 33.333)
        (35, 37.14)
        (40, 35)
        (50, 26)
        (60, 25)
        (70, 24.28)
        (80, 25)
        (90, 26.66677)
        (100, 23)
    };
      \addplot plot coordinates {
        (2,     50)
        (3,    66.6667)
        (4,    75)
        (5,   80)
        (6,   83.333)
        (7,   85.71)
        (8,   87.5)
        (9,   77.7778)
        (10,  70)
        (15,  60)
        (20,  40)
        (25,  40)
        (30, 33.333)
        (35, 37.14)
        (40, 37.5)
        (50, 28)
        (60, 25)
        (70, 25.714)
        (80, 28.75)
        (90, 27.7778)
        (100, 24)
    };

 \addplot plot coordinates {
        (2,     50)
        (3,    66.6667)
        (4,    75)
        (5,   80)
        (6,   83.333)
        (7,   85.71)
        (8,   87.5)
        (9,   77.7778)
        (10,  70)
        (15,  60)
        (20,  45)
        (25,  40)
        (30, 33.333)
        (35, 37.14)
        (40, 37.5)
        (50, 30)
        (60, 25)
        (70, 25.714)
        (80, 28.75)
        (90, 27.7778)
        (100, 24)
    };

    \addplot plot coordinates {
        (2,     50)
        (3,    66.6667)
        (4,    75)
        (5,   80)
        (6,   83.333)
        (7,   85.71)
        (8,   87.5)
        (9,   77.7778)
        (10,  70)
        (15,  60)
        (20,  45)
        (25,  40)
        (30, 36.6667)
        (35, 37.14)
        (40, 40)
        (50, 32)
        (60, 26.666)
        (70, 28.571)
        (80, 31.25)
        (90, 30)
        (100, 27)
    };

    \legend{$th=0.005$\\$th= 0.025$\\$th=0.05$\\$th=0.075$\\$th=0.10$\\}
 \end{axis}
\end{tikzpicture}}
\scalebox{0.45}{
\begin{tikzpicture}
    \begin{axis}[
        title={\textbf{Random}},
        xlabel=\textsc{Number of Clusters},
        ylabel= \textsc{Enrichment Score},
        xmin=0,xmax=100,ymin=0,ymax=100,
        legend style={anchor=north east}
    ]

         \addplot plot coordinates {
        (2,     100)
        (3,    100)
        (4,    100)
        (5,   100)
        (6,   100)
        (7,   100)
        (8,   100)
        (9,   100)
        (10,  100)
        (15,  100)
        (20,  100)
        (25,  100)
        (30, 86.666)
        (40, 37.5)
        (50, 36)
        (60, 11.6667)
        (70, 10)
        (80, 7.5)
        (90, 2.22)
        (100, 5)
    };
         \addplot plot coordinates {
        (2,     100)
        (3,    100)
        (4,    100)
        (5,   100)
        (6,   100)
        (7,   100)
        (8,   100)
        (9,   100)
        (10,  100)
        (15,  100)
        (20,  100)
        (25,  100)
        (30, 93.333)
        (40, 67.5)
        (50, 48)
        (60, 40)
        (70, 25.714)
        (80, 20)
        (90, 14.444)
        (100, 10)
    };
      \addplot plot coordinates {
        (2,     100)
        (3,    100)
        (4,    100)
        (5,   100)
        (6,   100)
        (7,   100)
        (8,   100)
        (9,   100)
        (10,  100)
        (15,  100)
        (20,  100)
        (25,  100)
        (30, 93.333)
        (40, 82.5)
        (50, 64)
        (60, 50)
        (70, 31.428)
        (80, 36.25)
        (90, 24.444)
        (100, 15)
    };

    \addplot plot coordinates {
        (2,     100)
        (3,    100)
        (4,    100)
        (5,   100)
        (6,   100)
        (7,   100)
        (8,   100)
        (9,   100)
        (10,  100)
        (15,  100)
        (20,  100)
        (25,  100)
        (30, 96.6667)
        (40, 85)
        (50, 64)
        (60, 56.6667)
        (70, 41.429)
        (80, 38.75)
        (90, 30)
        (100, 23)
    };

       \addplot plot coordinates {
        (2,     100)
        (3,    100)
        (4,    100)
        (5,   100)
        (6,   100)
        (7,   100)
        (8,   100)
        (9,   100)
        (10,  100)
        (15,  100)
        (20,  100)
        (25,  100)
        (30, 96.6667)
        (40, 85)
        (50, 64)
        (60, 56.667)
        (70, 41.4286)
        (80, 38.75)
        (90, 30)
        (100, 23)
    };

    \legend{$th=0.005$\\$th= 0.025$\\$th=0.05$\\$th=0.075$\\$th=0.10$\\}
 \end{axis}
\end{tikzpicture}}

\caption{\textbf{Evaluation of the clustering performance for different Enrichment Threshold values.}
      LP-Stability (upper left), CorEx (upper right), K-Means (lower left), Random (lower right). The figure presents the percentage of the enriched clusters for the threshold values of $0.005$, $0.025$, $0.05$, $0.075$, $0.1$ and using Kendall's correlation-based distance. The higher differences in the enrichment thresholds are reported from the Random Clustering when the number of clusters is relatively high. For the rest of the algorithms and especially LP-Stability, the different thresholds only slightly impact the reported results.}
\label{fig:gene_enrichment}
\end{figure}

\section{Dataset}
In this study, we based our experiments on The Cancer Genome Atlas (TCGA) dataset~\cite{grossman2016}. TCGA contains a comprehensive dataset including several data types such as DNA copy number, DNA methylation, mRNA expression, miRNA expression, protein expression, and somatic point mutation. 
We focused our study on tumor types relevant for radiotherapy and/or immunotherapy. For the gene clustering part, our dataset consists of \textbf{4615} samples (Table \ref{tab:tumors-locations} second column). 
In particular, we investigated the following types of tumors, namely: Urothelial Bladder Carcinoma (BLCA), Breast Invasive Carcinoma (BRCA), Cervical Squamous Cell Carcinoma and Endocervical Adenocarcinoma (CESC), Glioblastoma Multiforme (GBM), Head and Neck Squamous Cell Carcinoma (HNSC),  Liver Hepatocellular Carcinoma (LIHC), Rectum Adenocarcinoma (READ), Lung adenocarcinoma (LUAD), Lung Squamous Cell Carcinoma (LUSC) and Ovarian Cancer (OV). For each sample, we had the RNA-seq reads of \textbf{20 365} genes processed using normalized RNA-seq by Expectation-Maximization (RSEM)~\cite{Li2011}.

 Several articles as~\cite{Salem2017} consider the challenging and important task of generating biomarkers for distinguishing tumor and subtumor types. In this study, we also focus on this task 
basing our experiments on the cohort presented in~\cite{thorsson2018} by selecting samples from the $10$ locations used for the gene signature. This cohort consists of $3653$ samples (Table~\ref{tab:tumors-locations} third and fourth columns). 
For the tumor subtypes characterisation, we focused on subtypes that had more than $50\times n\_subtype$ samples. At the end, $5$ different tumor types namely the BRCA, HNSC, LIHC, READ and OV have been used for subtypes classification.

\section{Results and Discussion}
 This study has been designed upon three pivotal complementary aspects. The first one relates to the genes clustering performance to assess the definition of the signature regarding both a mathematical (DI) and a biological (ES) metric (section~\ref{section:res_gene_clust}). The second evaluates the ability of the signature to relevantly separate the different tumor samples in an unbiased manner in particular through sample clustering (section~\ref{section:unsup}). The third aspect characterizes the tissue specificity of the signature thanks to classification tasks on tumor types and subtypes (section~\ref{section:classification}). To better estimate the results obtained, comparisons with referential clustering methods and gene signatures are performed throughout the different evaluations. A global comparison with all references over all metrics is provided in section~\ref{section:compa}.

\subsection{Results on Clustering Gene Data}
\label{section:res_gene_clust}
The obtained clusters were evaluated using both mathematical and biological evaluation criteria.
Starting with the biological criteria, Fig.~\ref{fig:gene_enrichment} presents a comparison of the different ES per algorithm for different threshold ($th$) values. We observed that for the different clustering methods the threshold does not significantly change the behavior of the ES, indicating a strong statistical significance for the clusters. But, it is not the case for the random signature on which for a number of clusters higher than $30$ one can observe an important disparity between the different $th$ of the ES. For the rest of the study, we will use $th=0.005$.

\begin{figure*}[t!]
\centering
\scalebox{0.58}{
\begin{tikzpicture}
    \begin{axis}[
        title={\textbf{Random: Dunn's Index}},
        xlabel=\textsc{Number of Clusters},
        ylabel= \textsc{Dunn's Index},
        xmin=0,xmax=100,ymin=0,ymax=50,
        legend style={anchor=north east}
    ]
        \addplot plot coordinates {
        (2,     0.000986)
        (3,    0.000986)
        (4,    0.000986)
        (5,   0.000986)
        (6,   0.000986)
        (7,  0.000986)
        (8,  0.000986)
        (9,  0.000986)
        (10,  0.000986)
        (15,  0.000986)
        (20,  0.000986)
        (25,  0.000986)
        (30,  0.000986)
        (40,  0.000986)
        (50,  0.000986)
        (60,  0.000986)
        (70,  0.000986)
        (80,  0.000986)
        (90,  0.000986)
        (100,  0.000986)
    };
     \addplot plot coordinates {
        (2,     1.01)
        (3,    1.01)
        (4,    1.01)
        (5,   1.01)
        (6,   1.01)
        (7,   1.01)
        (8,  1.01)
        (9,  1.01)
        (10,  1.01)
        (15,  1.01)
        (20,  1.01)
        (25,  1.01)
        (30, 1.01)
        (40,  1.01)
        (50, 1.01)
        (60,  1.01)
        (70,  1.01)
        (80, 1.01)
        (90,  1.01)
        (100,  1.01)
    };
    \addplot plot coordinates {
        (2, 20.2)
        (3,    16.7)
        (4,    13.5)
        (5,   13.5)
        (6,   13.5)
        (7,   13.5)
        (8,  16.7)
        (9,  13.51)
        (10,  13.51)
        (15,  13.51)
        (20,  13.51)
        (25,  13.51)
        (30, 13.51)
        (40,  13.51)
        (50, 13.51)
        (60,  13.51)
        (70,  13.51)
        (80, 13.51)
        (90,  13.51)
        (100,  13.51)
    };
      \addplot plot coordinates {
        (2,     36.4)
        (3,    36.4)
        (4,    33.2)
        (5,   32.6)
        (6,   13.8)
        (7,   13.8)
        (8,  32.6)
        (9,  32.6)
        (10,  13.8)
        (15,  13.8)
        (20,  13.8)
        (25,  13.8)
        (30,  13.8)
        (40,  13.8)
        (50,  13.8)
        (60,  13.8)
        (70,  13.8)
        (80,  13.8)
        (90,  13.8)
        (100,  13.8)
    };
\addplot plot coordinates {
        (2,     0.000986)
        (3,    0.000986)
        (4,    0.000986)
        (5,   0.000986)
        (6,   0.000986)
        (7,  0.000986)
        (8,  0.000986)
        (9,  0.000986)
        (10,  0.000986)
        (15,  0.000986)
        (20,  0.000986)
        (25,  0.000986)
        (30,  0.000986)
        (40,  0.000986)
        (50,  0.000986)
        (60,  0.000986)
        (70,  0.000986)
        (80,  0.000986)
        (90,  0.000986)
        (100,  0.000986)
    };
    \addplot plot coordinates {
        (2,     0.000986)
        (3,    0.000986)
        (4,    0.000986)
        (5,   0.000986)
        (6,   0.000986)
        (7,  0.000986)
        (8,  0.000986)
        (9,  0.000986)
        (10,  0.000986)
        (15,  0.000986)
        (20,  0.000986)
        (25,  0.000986)
        (30,  0.000986)
        (40,  0.000986)
        (50,  0.000986)
        (60,  0.000986)
        (70,  0.000986)
        (80,  0.000986)
        (90,  0.000986)
        (100,  0.000986)
    };
 \end{axis}
\end{tikzpicture}}
\scalebox{0.58}{
\begin{tikzpicture}
    \begin{axis}[
        title={\textbf{LP-Stability: Dunn's Index}},
        xlabel=\textsc{Number of Clusters},
        ylabel= \textsc{Dunn's Index},
        xmin=0,xmax=100,ymin=0,ymax=50,
        legend style={anchor=north west}
    ]
        \addplot plot coordinates {
        (7,     0.000986)
        (7,    0.000986)
        (10,    0.000986)
        (16,   0.000986)
        (24,   0.000986)
        (35,  0.000986)
    };
     \addplot plot coordinates {
        (24,     1.37)
        (34,    5.32)
        (34,    5.32)
        (39,   5.32)
        (39,   5.32)
        (47,   5.32)
        (50,  5.32)
        (52,  5.32)
        (59,  5.32)
        (67,  5.32)
        (80,  5.32)
        (89,  5.32)
        (110, 5.32)
    };
    \addplot plot coordinates {
        (31,  20.3)
        (33,  20.3)
        (34,  20.3)
        (41, 20.3)
        (43,  20.3)
        (46, 20.3)
        (52,  20.3)
        (60,  20.3)
        (66, 20.3)
        (82,  20.3)
        (98,  19.5)
    };
      \addplot plot coordinates {
        (22,     40.3)
        (27,    40.6)
        (27,    40.6)
        (29,   40.6)
        (31,   39.9)
        (38,   36.8)
        (43,  37.2)
        (44,  37.2)
        (50,  37.5)
        (60,  37.5)
        (72,  37.6)
        (98,  36.8)

    };
\addplot plot coordinates {
        (15,     5)
        (35,    1)
        (36,    1)
        (36,   1)
        (42,   1.3)
        (57,  2)
        (57, 2)
        (59,  2)
        (60,  2)
        (75,  2)
    };
    \addplot plot coordinates {
        (15,     4.72)
        (35,    1)
        (37,    1)
        (37,   1)
        (38,   1)
        (40,  1)
        (53,  1)
        (55,  1)
        (56,  1)
        (64,  1)
        (66,  1)
        (82,  1)
        (97,  1)

    };
 \end{axis}
\end{tikzpicture}}
\scalebox{0.58}{
\begin{tikzpicture}
    \begin{axis}[
        title={\textbf{LP-Stability: Enrichment Score}},
        xlabel=\textsc{Number of Clusters},
        ylabel= \textsc{Dunn's Index},
        xmin=0,xmax=100,ymin=0,ymax=105,
        legend style={at={(1.05,0.8)},anchor=north west}
    ]
        \addplot plot coordinates {
        (7,     85.7)
        (7,    100)
        (10,    90)
        (16,   93.7)
        (24,   87.5)
        (35,  26)
    };
     \addplot plot coordinates {
        (24,     100)
        (34,    100)
        (34,    100)
        (39,   100)
        (39,   100)
        (47,   100)
        (50,  100)
        (52,  100)
        (59,  100)
        (67,  98.5)
        (80,  96.25)
        (89,  95.5)
        (110, 94.54)
    };
    \addplot plot coordinates {
        (31,  100)
        (33,  100)
        (34,  97)
        (41, 97.56)
        (43,  97.67)
        (46, 97.82)
        (52,  98)
        (60,  98.33)
        (66, 98.48)
        (82,  97.56)
        (98,  96)
    };
      \addplot plot coordinates {
        (22,     100)
        (27,    100)
        (27,    100)
        (29,   100)
        (31,   100)
        (38,   100)
        (43,  97.7)
        (44,  97.72)
        (50,  98)
        (60,  98.3)
        (72,  98.67)
        (98,  96)

    };
\addplot plot coordinates {
        (15,     5)
        (35,    1)
        (36,    1)
        (36,   1)
        (42,   1.3)
        (57,  2)
        (57, 2)
        (59,  2)
        (60,  2)
        (75,  2)
    };
    \addplot plot coordinates {
        (15,     4.72)
        (35,    1)
        (37,    1)
        (37,   1)
        (38,   1)
        (40,  1)
        (53,  1)
        (55,  1)
        (56,  1)
        (64,  1)
        (66,  1)
        (82,  1)
        (97,  1)

    };
    \legend{$Kullback-Leibler$\\$Pearson's$\\$Spearman's$\\$Kendall's$\\$Euclidean$\\$Cosine$\\}
 \end{axis}
\end{tikzpicture}}
\caption{\textbf{Evaluation of the clustering performance for different distances.} The performance of the different distances are presented for both Random (left) in terms of Dunn's Index and LP-Stability clustering (middle and right) in terms of Dunn's Index and Enrichment Score. Only DI results are presented for Random as ES computation on a same clustering is not influenced by the distance used. Both ES and DI are presented in percentages in terms of the number of clusters. The figure highlights the superiority of the   correlation-based distances and in particular the one reported by Kendall's for both mathematical and biological aspects.}
\label{fig:gene_distances}
\end{figure*}
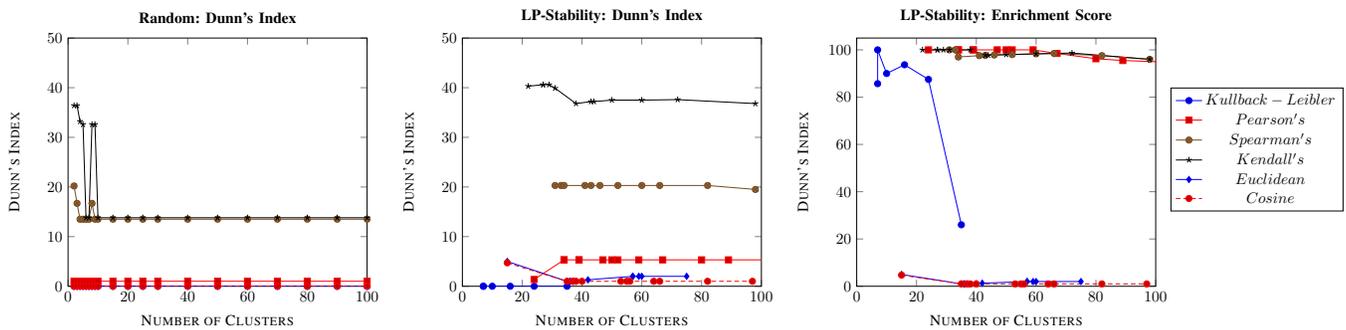


To select the best distance per method we used the DI metric. In Fig.~\ref{fig:gene_distances} one can observe the influence of the distance with respect to the number of clusters for the random and LP-stability methods. Compared with random clustering one can observe the bias that each distance introduces for the DI score. In particular, with correlation-based distances the reported DI scores are on average $10$ times higher. Thus, to tackle this problem of bias, for our comparisons, we will refer to a clustering difference in DI scores with the corresponding random clustering for the same number of clusters and distance.
Based on our experiments we also noticed that the different distances greatly affects the performance of the clustering algorithm, with the correlation-based distances (especially the Kendall's correlation) reporting in general higher performances.
To ensure the biological meaning of the clusters, we also report the performance of the different distances for ES. Once again the superiority of correlation-based distances both in terms of performance and stability is indicated. Besides, only the Euclidean distance does not reach the maximal value of $100\%$. This is due to the unbalanced clusters that Euclidean distance favors, leading to very small clusters that are less likely to be enriched. For the rest of the paper we selected Kendall's correlation-based distance when reporting LP-Stability and Random Clustering performances.


In Table~\ref{tab:bests}, we summarize the performance of LP-Stability in comparison to other algorithms based on both ES and DI scores together with the reported number of clusters. Additional information about the average Enrichment and the average computational time per algorithm is also provided in the table. The best performance of DI is achieved with the LP-Stability and the Kendall's Correlation-based distance. Moreover, even if almost all the methods, except K-Means, reached an Enrichment Score of $100\%$, LP-Stability still reports the highest average Enrichment, with $96\%$ while CorEx reaches only $71\%$.
Another interesting point from this analysis is the indication of the optimal number of clusters per algorithm. Only LP-Stability reports its best value with more than $25$ clusters while the rest of the algorithms have their best performance with less than $7$ clusters and even $2$ clusters only if we consider DI alone. This might seem to be an argument in favor of the other algorithms as they are able to define a more compact signature. However, such a low number of clusters highlights failure on characterizing a clustering structure as they favor a disposition where genes are grouped altogether. This is also indicated by the low average ES and DI scores. Computational time study was presented in~\cite{battistella2019} and can be found in Supplementary Material.

 \begin{table*}[!t]
\caption{Comparison of the different evaluated algorithms in terms of PPI Enrichment Score (ES) with a threshold of $0.005$, Dunn's Index (DI), Average ES and computational time. LP-Stability algorithm outperforms the rest of the algorithms reporting highest DI and Average ES score and the lowest computational time.}
\label{tab:bests}
\centering
\begin{tabular}{|c|c|c|c|c|c|c|c|c|c|}
\hline
\multirow{2}{*}{Method}   & \multicolumn{3}{c|}{Best ES} & \multicolumn{3}{c|}{Best   DI} & \multirow{2}{*}{Average ES (\%)} & \multirow{2}{*}{Average DI (\%)} & \multirow{2}{*}{Time} \\ \cline{2-7}
                          & ES (\%) & DI (\%) & Clusters & ES (\%)  & DI (\%)  & Clusters &                                  &                                  &                       \\ \hline
Random                    & 100     & 36      & 2        & 100      & 36       & 2        & 54                               & 19.8                             & -                     \\ \hline
K-Means (Euclidean)       & 85.7    & 2.5     & 7        & 50       & 15.6     & 5        & 37                               & 1.2                              & 3h                    \\ \hline
CorEx (Total Correlation) & 100     & 2.4     & 5        & 100      & 2.4      & 5        & 71                               & 0.6                              & \textgreater{}5 days  \\ \hline
LP-Stability (Kendall’s)  & 100     & 40.6    & 27       & 100      & 40.6     & 27       & 96                               & 38.5                             & 1.5h                  \\ \hline
\end{tabular}
\end{table*}

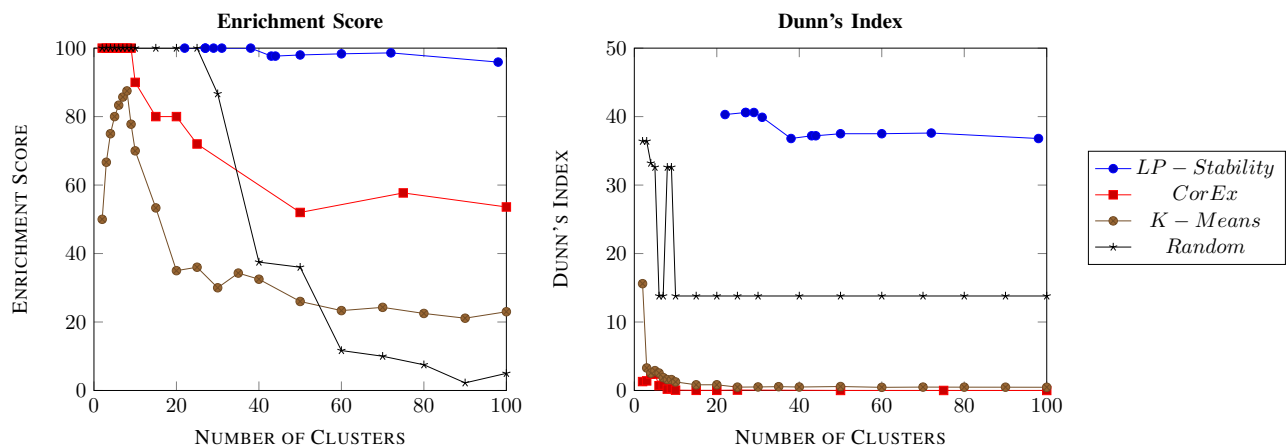
\begin{figure*}[!t]
\centering
\scalebox{0.8}{
\begin{tikzpicture}
    \begin{axis}[
        title={\textbf{Enrichment Score}},
        xlabel=\textsc{Number of Clusters},
        ylabel= \textsc{Enrichment Score},
        xmin=0,xmax=100,ymin=0,ymax=100,
        legend style={at={(0.03,0.02)},anchor=south west}
    ]

    \addplot plot coordinates {
        (22,     100)
        (27,    100)
        (27,    100)
        (29,   100)
        (31,   100)
        (38,   100)
        (43,  97.6744186)
        (44,  97.6744186)
        (50,  98)
        (60,  98.33333333)
        (72,  98.61111111)
        (98,  95.91836735)
    };

      \addplot plot coordinates {
        (2,     100)
        (3,    100)
        (4,    100)
        (5,   100)
        (6,   100)
        (7,   100)
        (8,  100)
        (9,  100)
        (10,  90)
        (15,  80)
        (20,  80)
        (25,  72)
        (50, 52)
        (75, 57.703)
        (100, 53.6)
    };

    \addplot plot coordinates {
        (2,     50)
        (3,    66.6667)
        (4,    75)
        (5,   80)
        (6,   83.333)
        (7,   85.71)
        (8,   87.5)
        (9,   77.7778)
        (10,  70)
        (15,  53.333)
        (20,  35)
        (25,  36)
        (30, 30)
        (35, 34.29)
        (40, 32.5)
        (50, 26)
        (60, 23.333)
        (70, 24.28)
        (80, 22.5)
        (90, 21.111)
        (100, 23)
    };

 \addplot plot coordinates {
        (2,     100)
        (3,    100)
        (4,    100)
        (5,   100)
        (6,   100)
        (7,   100)
        (8,   100)
        (9,   100)
        (10,  100)
        (15,  100)
        (20,  100)
        (25,  100)
        (30, 86.666)
        (40, 37.5)
        (50, 36)
        (60, 11.6667)
        (70, 10)
        (80, 7.5)
        (90, 2.22)
        (100, 5)
    };
 \end{axis}
\end{tikzpicture}}
\scalebox{0.8}{
\begin{tikzpicture}
    \begin{axis}[
        title={\textbf{Dunn's Index}},
        xlabel=\textsc{Number of Clusters},
        ylabel= \textsc{Dunn's Index},
        xmin=0,xmax=100,ymin=0,ymax=50,
        legend style={at={(1.1,0.7)},anchor=north west}
    ]
    \addplot plot coordinates {
        (22,     40.3)
        (27,    40.6)
        (27,    40.6)
        (29,   40.6)
        (31,   39.9)
        (38,   36.8)
        (43,  37.2)
        (44,  37.2)
        (50,  37.5)
        (60,  37.5)
        (72,  37.6)
        (98,  36.8)

    };
     \addplot plot coordinates {
        (2,     1.3)
        (3,    1.4)
        (4,    2.4)
        (5,   2.4)
        (6,   0.7)
        (7,   0.8)
        (8,  0.2)
        (9,  0.3)
        (10,  0.05)
        (15,  0.02)
        (20,  0.02)
        (25,  0.06)
        (50, 0.005)
        (75, 0.003)
        (100, 0.003)
    };

    \addplot plot coordinates {
        (2,     15.6)
        (3,    3.29)
        (4,    2.54)
        (5,   2.90)
        (6,   2.55)
        (7,   1.89)
        (8,   1.59)
        (9,   1.59)
        (10,  1.25)
        (15,  0.83)
        (20,  0.834)
        (25,  0.489)
        (30, 0.516)
        (35, 0.555)
        (40, 0.507)
        (50, 0.577)
        (60, 0.461)
        (70, 0.495)
        (80, 0.485)
        (90, 0.48)
        (100, 0.47)
    };
      \addplot plot coordinates {
        (2,     36.4)
        (3,    36.4)
        (4,    33.2)
        (5,   32.6)
        (6,   13.8)
        (7,   13.8)
        (8,  32.6)
        (9,  32.6)
        (10,  13.8)
        (15,  13.8)
        (20,  13.8)
        (25,  13.8)
        (30,  13.8)
        (40,  13.8)
        (50,  13.8)
        (60,  13.8)
        (70,  13.8)
        (80,  13.8)
        (90,  13.8)
        (100,  13.8)
    };
      
     \legend{$LP-Stability$\\$CorEx$\\$K-Means$\\$Random$\\}
 \end{axis}
\end{tikzpicture}}
\caption{\textbf{Evaluation of the different clustering algorithms.}
  For the different evaluated algorithms the ES and the DI are presented in terms of number of clusters and using Kendall's correlation-based distance. For both metrics, LP-Stability reports the highest and more stable values. Moreover, the rest of the algorithms tends to report their higher scores for a very small number of clusters (often $2$), indicating their failure to discover clustering structures.}
\label{fig:graph_comparison}
\end{figure*}

A thorough comparison of the different algorithms for a different number of clusters is presented in Fig.~\ref{fig:graph_comparison}. For both DI and ES the superiority of the proposed LP-Stability in comparison to the other algorithms can be observed both in terms of stability for a varying number of clusters and performance. The reported results indicate that the proposed method can generate clusters that are both mathematically and biologically meaningful. Moreover, one can observe that for Random Clustering, the reported enrichment is very high, however dropping dramatically for more than $30$ clusters, while the DI is really low for all the cases. This highlights the need to study both the mathematical performance and the stability of the biological score as ES alone would not give significant results. 

\subsection{Unsupervised Signature Assessment}
\label{section:unsup}
\subsubsection{Signature Selection}

 The signature was selected using the method detailed in section~\ref{section:ss} on the clustering presenting the highest DI among clusterings having best ES. However, due to the relatively low number of genes for signatures based on K-Means, CorEx or Random Clustering, the sample clusterings with those signatures gave quite irrelevant intermixed tumor types (Fig.1 in Supplementary). To deal with this and for comparison reasons, we used for all these algorithms the gene signatures produced with $25$ and $30$ genes and in the following, when referring to CorEx and K-Means signatures we will refer to those signatures.
 To facilitate future studies leveraging this work, we provide in additional material the clustering of genes and samples.

Regarding the evaluation of the enriched biological processes for the different signatures, we found that LP-Stability signature ($27$ genes with Kendall's correlation-based distance) does not present any redundancy in the biological processes, in contrast to the K-Means ($30$ genes with Euclidean distance) which presents several hundred of enriched biological processes. Moreover, CorEx signature ($30$ genes with Total Correlation)  presents a low biological redundancy with only phototransduction process being enriched. 

Our proposed gene signature using LP-Stability is composed by $27$ genes. 
Globally these genes, which descriptions are detailed in Supplementary Materials, are related to cell development and cell cycle (CD53, NCAPH, GNA15, GADD45GIP1, CD302, NCAPH, YEATS2), DNA transcription (HSFX1, CCDC30, MATR3, ASH1L, ANKRD30A, GSX1), gene expression (ZNF767, C1orf159, RPS8, ZEB2), DNA repair (RIF1), antigen recognition (ZNF767), apoptosis (C3P1, CLIP3), mRNA splicing (SNRPG). We also have many genes specific to cancer or having a major impact on cancer (CD53, ANKRD30A, ZEB2, ADNP, SFTA3, ACBD4). All these processes are highly important and significant for cancer. We also report in Supplementary Materials for each gene the main tissues they are overexpressed using GTEx portal. Finally, even if we have many genes related to specific tissue types such as brain, blood lymphocytes, liver or gynecologic tissues, the overall profiles of each gene are unique.
\subsubsection{Sample Clustering: Discovery Power}
\begin{figure*}[!t]
\centering
\includegraphics[scale=0.5]{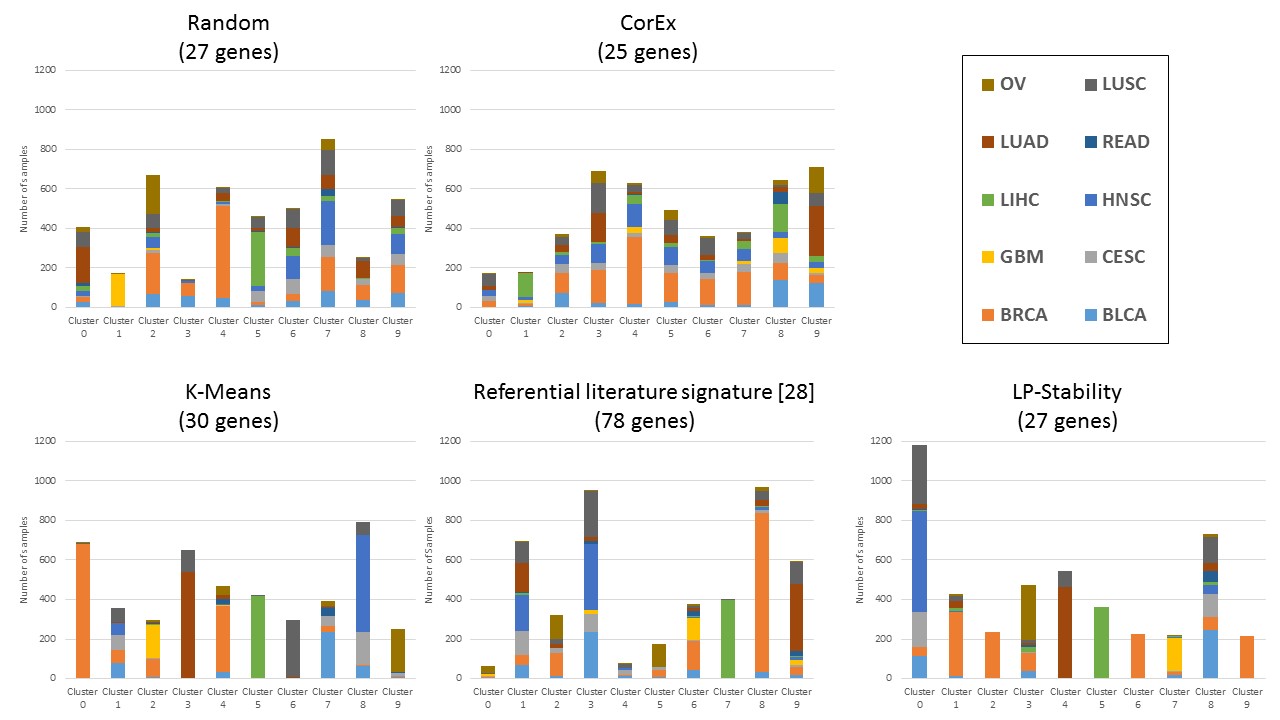}
\caption{\textbf{Gene Signature Assessment via tumor distribution analysis across 10 sample clusters generated in the different signatures feature space.} The graph presents the distribution of the different tumors for Random Clustering signature ($27$ genes)
, CorEx  
, K-Means  
, a referential gene signature~\cite{thorsson2018} 
and LP-Stability.
The distribution of tumors for Random and CorEx algorithms is quite intermixed without a lot of associations between the tumor types while K-Means, referential and LP-Stability signatures seem to favor some good tumor associations.}
\label{fig:sample_clustering}
\end{figure*}

The predictive powers of the best signature per algorithm together with the random signatures, and the signature presented in~\cite{thorsson2018}, were further assessed by measuring their ability to separate $10$ different tumor types (Table~\ref{tab:tumors-locations}) in a completely unsupervised manner,   through sample clustering. In Fig.~\ref{fig:sample_clustering}, the results for the LP-Stability (with $27$ clusters, ES $100\%$ and DI $40.6\%$ using Kendall's correlation-based distance) signature, K-Means (with $30$ genes, ES of $30\%$ and a DI of $0.52\%$ using Euclidean distance), CorEx (with $25$ genes, ES $72 \%$ and DI $0.06 \%$ using Total Correlation), Random Clustering signature (with $27$ genes, ES $86.6\%$ and DI of $13.8\%$ using Kendall's correlation-based distance) and the signature from~\cite{thorsson2018} are presented. One can observe that CorEx and Random signatures fail to properly separate the tumor types and for this reason for the rest of the section we present a detailed comparison of the K-Means and LP-Stability signatures only. A thorough analysis of the clusterings of Fig.~\ref{fig:sample_clustering} is provided in ~\cite{battistella2019} and in Supplementary Materials and shows evidence validating the relevance of the approach presented here.


Another interesting point is that the distance used greatly affects the distribution of the different tumor types for the clustering. This proves the importance of the distance selection in combination with the selected algorithm. Based on our experiments, we noticed that Spearman's and Kendall's correlations provide the best sample clustering for all the algorithms. In particular, Spearman's correlation tends to better separate the different tumors into different clusters, while the Kendall's seems to generate clusters that groups tumor-related samples (Fig.2 in Supplementary Materials).

To compare our results with other methods in the literature, we assessed our gene signature against a knowledge-based signature of $78$ genes that has been proven to be appropriate for determining immune related sample clusters~\cite{thorsson2018}. The obtained tumor distribution is presented in Fig.~\ref{fig:sample_clustering}, reporting quite intermixed associations. Again, LIHC and BRCA are separated properly while the rest of the tumors are clustered in unrelated groups. This comparison indicates the need for compact signatures, highlighting at the same time the difficulty of capturing the full genome information as well as the need for an automatically computed signature to avoid redundancy and information loss.


\subsubsection{Gene Screening Analysis}
 \label{section:screening}
Screening analysis aims to identify the significant genes for each cluster which are then used to determine the enriched biological processes per cluster. To determine those significant genes, we used the SAM method to look for genes that are expressed differently for the samples of one tumor type in a cluster compared to the other samples of the same tumor type (see section~\ref{section:DP} for more details). We will refer to those genes as differentially expressed genes in the remainder of the article. Besides, the SAM method scores allows to determine significant genes, in the following, we will refer in particular to the most significant genes for a cluster or a tumor type in a cluster for the genes reaching the highest SAM scores.  This method, allows to check if the tumor types of a given cluster share genes related to similar biological processes, highlighting the biological relevance of the cluster. Meanwhile, this method enables us to verify the relevance of the distribution of the same tumor types into different clusters by checking the absence of similar biological processes.
In particular, a summary of the analysis for our proposed signature's sample clustering is presented in Table~\ref{tab:gene_screening}. In this section, we provide a detailed analysis per tumor type for each cluster for both K-Means and LP-Stability selected signatures. 

\begin{table*}[!t]
\caption{Analysis of the biological pathways and most significant genes per cluster for the sample clustering performed using our proposed signature of $27$ genes via LP-Stability algorithm and Kendall's correlation-based distance. The table highlights the separation between inflamed and non-inflamed tumors and the identification of well-known cancer subtypes such as BRCA.}
\label{tab:gene_screening}
\centering
\scalebox{0.95}{
\begin{tabular}{|c| c| c| c| c| c|}
\hline
\multirow{2}{*}{\textbf{Clusters}} & \textbf{Significant} & \textbf{Tumor} & \textbf{Validated Role of} \textbf{the Gene}  & \textbf{Biological Pathways} & \textbf{Key Feature} \\
& \textbf{Genes} & \textbf{Type} & \textbf{in Tumor Type} & & \textbf{of the cluster}\\\hline

 \multirow{4}{*}{Cluster 0} & IL4R & HNSC & Treatment target & Immune and defense response &\multirow{2}{*}{An inflamed solid } \\ \cline{2-4}
 & GNA15 & LUSC &  Lung cancers treatment & Regulation of cell proliferation& \multirow{2}{*}{tumors cluster}\\ \cline{2-4}
 & \multirow{2}{*}{KRT5} & \multirow{2}{*}{CESC} & Biomarker distinguishing adenocarcinomas  & \multirow{2}{*}{Interferon-gamma mediated process}& \\
  & & &  from squamous cell carcinomas~\cite{Xiao2017} & &\\\hline
 \multirow{2}{*}{Cluster 1} & \multirow{2}{*}{FPR3} & \multirow{2}{*}{BRCA} & \multirow{2}{*}{Immune inflammation related~\cite{li2017}}  &  \multirow{2}{*}{Immune response related pathways} & An inflamed BRCA\\
 &&&&& tumors cluster \\ \hline
 \multirow{2}{*}{Cluster 2} & \multirow{2}{*}{TTC28} & \multirow{2}{*}{BRCA} & \multirow{2}{*}{Related to basal breast cancer risk~\cite{Hamdi2016}} & \multirow{2}{*}{Basal plasma membrane} &  A basal BRCA\\
 &&&&& tumors cluster\\ \hline
 \multirow{5}{*}{Cluster 3} &  \multirow{3}{*}{NDUFB10} &  \multirow{3}{*}{BRCA} & Related to breast~\cite{Zhang2015} and ovarian~\cite{PermuthWey2011}  & \multirow{2}{*}{Mitochondrial complexes and}  & \multirow{5}{*}{A gynecologic tumors}\\ 
  &&&  cancers, poor prognosis for  & \multirow{3}{*}{cells organization related processes} &  \\
  &&&  esophageal lineage and LIHC &&\\\cline{2-4}
 &  \multirow{2}{*}{SLC39A6} &  \multirow{2}{*}{OV} & Poor prognosis for  & & cluster linked to LIHC\\
 &  &&esophageal lineage and LIHC & & \\\hline
 \multirow{2}{*}{Cluster 4} & SFTA3 & LUAD & Related to LUAD and LUSC~\cite{Schicht2014, Xiao2017}  & Pathways of immune response & \multirow{2}{*}{An inflamed lung tumors cluster}\\ \cline{2-4}
 & NAPSA & LUSC & Related to LUAD  & Surfactant homeostasis & \\\hline
 \multirow{2}{*}{Cluster 5} &  & \multirow{2}{*}{LIHC}& &  & A pure complete LIHC\\
                            &&&&& tumor cluster\\ \hline
 \multirow{2}{*}{Cluster 6} &  \multirow{2}{*}{FOXA1} &  \multirow{2}{*}{BRCA} & Related to Breast   &  \multirow{2}{*}{Metabolic processes} & A luminal BRCA \\ 
 &&& Luminal cancer~\cite{Cappelletti2017} && tumors cluster\\\hline
 \multirow{4}{*}{Cluster 7} & & GBM & Complete GBM cluster & Response to stimulus, &\\\cline{2-4}
 & ANGPTL5 & BRCA & Angiopoietin-like protein family  &  \multirow{2}{*}{cardiovascularity,} & A GBM tumors cluster  \\ \cline{2-4}
 & \multirow{2}{*}{FERMT2} & \multirow{2}{*}{BLCA} & Related to various cancer  &   & with other tumors, all enriched\\
  & &&including breast ones & blood vessels related &  in cardiovascularity pathways\\\hline
 \multirow{2}{*}{Cluster 8} & UQCRH & BRCA & Mitochondrial Hinge protein related~\cite{Modena2003} & Metabolic processes & \multirow{2}{*}{Mis-splicing related tumors}\\ \cline{2-4}
 & AP1M2 & BLCA & Tyrosine-based signals  & general compound processes & \\\hline
 \multirow{2}{*}{Cluster 9} & \multirow{2}{*}{CIRBP} & \multirow{2}{*}{BRCA} & \multirow{2}{*}{Driver of many cancers}  & Related to  alternative splicing & \multirow{2}{*}{Alternative splicing related BRCA}\\ 
 &  & && processes and organelles & \\ \hline

\end{tabular}}
\end{table*}

Starting with the gene screening of LP-Stability, cluster $0$ is one of the most intermixed clusters. This cluster contains significant genes for different tumor types that are associated with immune, defense response and other inflammatory processes with strong enrichment. Among the most significant genes we can report IL4R for HNSC, this Interleukin is a treatment target for multiple cancers, GNA15 for LUSC which has been highlighted in lung cancer treatment or for CESC. KRT5, has been identified as a potential biomarker to distinguish adenocarcinomas to squamous cell carcinomas~\cite{Xiao2017}. Continuing our analysis with cluster $1$ which includes mainly BRCA samples, its most significant gene is FPR3. This gene seems to be related to immune inflammation and multiple cancers including breast~\cite{li2017}. For cluster $2$ which is mainly composed of BRCA samples, we identified that it is a basal BRCA cluster. Indeed, its most significant gene TTC28 is related to breast cancer and especially basal BRCA~\cite{Hamdi2016}. Besides, the cluster is enriched for basal plasma membrane. 
Cluster $3$ is also a mixed cluster mostly composed of BRCA and OV cancers. It seems to be related to mitochondrial complexes and organization. Its most significant gene is the NDUFB10 which is related to breast cancer patients~\cite{Zhang2015}, OV cancer~\cite{PermuthWey2011} and also correlated to a decreased viability in esophageal squamous lineage as LIHC. 
Cluster $4$ is a lung related cluster, composed of LUAD/LUSC samples. LUAD samples are related to immune response and LUSC samples to surfactant homeostasis which is linked to many lung diseases. The most significant gene for LUAD is SFTA3, a lung protein~\cite{Schicht2014} and a biomarker distinguishing LUAD and LUSC~\cite{Xiao2017}, whereas the most significant gene for LUSC samples is NAPSA that has been proven to be of relevance for LUAD tumors. 
Cluster $5$ mostly consists of LIHC samples, grouping all the LIHC samples in this cluster. Similarly cluster $7$ consists of all GBM tumors.
Thus, the screening process is not applicable for them as it compares samples from the same tumor type over different clusters.
Cluster $6$ is a luminal breast cancer cluster, related to metabolic processes which have already been studied in a breast cancer context~\cite{Schramm2010}. The most significant gene seems to be the FOXA1, a gene related to Estrogen-Receptor Positive Breast Cancer and Luminal Breast Carcinoma~\cite{Cappelletti2017}. 
Cluster $7$ is GBM tumors cluster. It is interesting to notice that next two dominant tumor types in the cluster, BRCA and BLCA, are related to cardiovascularity and blood vessels, their respective most significant genes are ANGPTL5 and FERMT2. The latter having been highlighted in GBM proliferation~\cite{Alshabi2019}. 
Cluster $8$ has no biological process linked to immune response, but presents a strong association to metabolic and structural processes. This group of processes has been found significant for BRCA~\cite{Read2018}. The most significant genes for this cluster are the AP1M2 for BLCA samples which interacts in tyrosine-based signals and has been considered in epithelial cells studies, the UQCRH  for BRCA a gene encoding mitochondrial Hinge protein that is important in soft tissue sarcomas and in particular in two cell lines of breast cancer and one of ovarian cancer~\cite{Modena2003}. 
Finally, cluster $9$ is a cluster with BRCA tumors, it has CIRP as its most significant gene, which is considered to be an oncogene in several cancers and in particular for BRCA. Cluster $9$ presents alternative splicing and coiled coil processes. 

This analysis highlights that each cluster is enriched in similar biological processes while the processes from different clusters are different. Moreover, it reveals that even if clusters $0$ and $8$ contain different tumor types, they present a homogeneity in their biological processes. Cluster $0$ is especially interesting as it contains inflamed tumor samples and cluster $8$ non-inflamed samples. These two clusters contain all the CESC samples, proving once more the relevance of the LP-Stability signature as they automatically and without any prior knowledge separate inflammatory and non-inflammatory CESC samples. This specific problem is an active field of research~\cite{Heeren2016}. Clusters $0$ and $8$ provide an even more valuable insight when studying the genes IFNG, STAT1, CCR5, CXCL9, CXCL10, CXCL11, IDO1, PRF1, GZMA, MHCII and HLA-DRA  highlighted in~ \cite{ayers2017} for their major role in immunotherapy. Indeed, for each tumor type in cluster $0$ all or most of these genes are differentially expressed which is not the case for cluster $8$, so proving the specificity and clinical relevance of the separation of these clusters. 

On a second level, we analyzed the distribution of the BRCA cancer samples on different clusters, examining its clinical relevance. We chose to highlight BRCA in this comparison, as it is the most represented tumor type and it presents a variety of subtypes. BRCA samples are distributed into clusters $1$, $2$, $3$, $6$, $8$ and $9$ using the LP-Stability signature, featuring the main molecular subtypes of BRCA. In particular, cluster $1$ contains immune inflammatory samples, cluster $2$ basal samples and cluster $6$ the luminal Estrogen-Receptor Positive samples. Additionally cluster $3$ is a gynecologic cluster with BRCA samples presenting relations to OV samples. Cluster $8$ features mis-plicing related tumors  which are strongly related to BRCA samples~\cite{Koedoot2019}. Cluster $9$ is marked by alternative splicing whose implications in cancers are well known and studied~\cite{singh2017}. It is also interesting to report that hallmarks genes BRCA1 and BRCA2 are positively and differentially expressed in luminal BRCA cluster $6$ which attests of an over-expression of these genes for cluster $6$. This observation is consistent with~\cite{mahmoud2017}, where BRCA1 and BRCA2 were more expressed in luminal BRCA samples as they are markers for good prognosis. Besides, these genes present an under-expression in cluster $3$ which is coherent as this mixed cluster groups BRCA and OV samples that are known to present a bad prognosis.

For comparison, we performed the same analysis with the sample clustering produced by the K-Means algorithm with the $30$ genes. After the analysis, we observed that this signature seems to be very specific for the BRCA tumors while reporting weaker relevance in the separation of other samples. Indeed, we can observe that the separation of BRCA samples is rather meaningful as in each cluster BRCA samples present rather relevant differentially expressed genes and enriched biological processes. However, the clusters are lacking homogeneity as the different tumor types of the clusters present unrelated differentially expressed genes and enriched biological processes. Besides, K-Means signature fails to properly characterize other tumor types. This issue might be explained by the over-representation of BRCA samples in our data set. Detailed analysis of the sample clustering using K-Means signature can be found in Supplementary Materials.   

Additionally, in order to indicate the significance of the distance used we considered different distances to perform the sample clustering. Our experiments confirmed that the distance that gave the best biologically relevant clusters was Spearman's correlation-based distance.
Moreover, after the screening analysis we observed that the differentially expressed genes are not necessarily the genes selected in the signatures. This observation indicates that the strength of our approach is to combine  genes that might not be the most informative taken individually but whose combination allows a good compact representation of the information brought by the whole genome for cancer tumors. It is also worth mentioning that LP-Stability signature correctly separates immune inflammatory samples from the others for all tumor types.



\subsubsection{Expression Power}
The expression power of our signature was further evaluated using the ARI, NMI, homogeneity, completeness and FMS metrics and compared with the rest of the signatures. We called the average of those scores the Expression Power of the signature and report it in Fig.\ref{fig:spider}. Detailed results for each score are provided in Supplementary Materials. For Random Clustering, we calculated the metrics on the average results of sample clustering designed from $10$ random signatures. Overall the performances of K-Means and LP-Stability are the best with the first outperform the second. Good performance of K-Means could be due to the good separation of BRCA clusters, the dominant tumor type in our dataset.


\subsection{Tumor Types/Subtypes Classification Tasks}
\label{section:classification}
 The predictive power of our proposed signature has been assessed in a supervised setting by classifying the samples according to their tumor and sub-tumor types. This experiment aims to evaluate the tissue-specific information captured by each signature. 
 In Table~\ref{tab:types}, we report the performance on training and test for each signature using the same classification strategy. Our experiments highlight that even random signatures with the relatively small number of $27$ genes reports good performance with a balanced accuracy of $84\%$. This proves that even a low number of genes are informative enough to perform a good separation of tumor types.  
However, our proposed signature reports the highest balanced accuracy reaching $92\%$ outperforming the referential signature~\cite{thorsson2018} which reached a balanced accuracy of $85\%$.

Regarding tumor subtypes classification, results averaged over all considered tumor types are provided in Table~\ref{tab:subtypes}. Our proposed method presents the highest performance with a balanced accuracy of $68\%$, outperforming the other algorithms by at least $9\%$. This task is quite challenging as we are using the same compact signature to characterize all the different tumor types at a fine molecular level. Considering the complexity of the task and the important number of different classes, results obtained with the proposed signature are very promising. Indeed, it is surpassing the random signatures average balanced accuracy by $11\%$ and the referential signature, devised on this specific dataset, by $9\%$.

\begin{table*}[!t]
\caption{\textbf{Tumor Types Classification} performance using the average performance of $10$ sets of randomly-selected genes of same size as the proposed signature, CorEx, K-Means, the referential~\cite{thorsson2018} and our proposed signatures.}
\label{tab:types}
\begin{center}
\begin{tabular}{|c|c|c|c|c|c|c|c|c|}
\hline
\multirow{2}{*}{Signature} & \multicolumn{2}{c|}{Balanced Accuracy (\%)} & \multicolumn{2}{c|}{Weighted Precision (\%)} & \multicolumn{2}{c|}{Weighted Sensitivity   (\%)} & \multicolumn{2}{c|}{Weighted Specificity   (\%)} \\ \cline{2-9} 
                           & Training             & Test                 & Training              & Test                 & Training                & Test                   & Training                & Test                   \\ \hline
Random                     & 96+/-5               & 84+/-2               & 95+/-5                & 87+/-3               & 94+/-7                  & 86+/-4                 & 99+/-1                  & 97+/-1                 \\ \hline
CorEx                      & 100                  & 85                   & 100                   & 90                   & 100                     & 91                     & 100                     & 98                     \\ \hline
K-Means                    & 100                  & 90                   & 100                   & 94                   & 100                     & 94                     & 100                     & 98                     \\ \hline
Referential {[}28{]}       & 100                  & 85                   & 100                   & 89                   & 100                     & 89                     & 100                     & 98                     \\ \hline
\textbf{Proposed}          & \textbf{99}          & \textbf{92}          & \textbf{99}           & \textbf{94}          & \textbf{98}             & \textbf{93}            & \textbf{100}            & \textbf{99}            \\ \hline
\end{tabular}
\end{center}

\end{table*}

\begin{table*}[!t]
\caption{\textbf{Tumor Subtypes Classification} performance using the average performance using $10$ sets of randomly-selected genes of same size as the proposed signature, CorEx, K-Means, the referential~\cite{thorsson2018} and our proposed signature. Only the $5$ types of tumors with more than $50 \times n\_subtypes$ samples were studied}
\label{tab:subtypes}
\begin{center}
\begin{tabular}{|c|c|c|c|c|c|c|c|c|}
\hline
\multirow{2}{*}{Signature} & \multicolumn{2}{c|}{Balanced Accuracy (\%)} & \multicolumn{2}{c|}{Weighted Precision (\%)} & \multicolumn{2}{c|}{Weighted Sensitivity   (\%)} & \multicolumn{2}{c|}{Weighted Specificity   (\%)} \\ \cline{2-9} 
                           & Training              & Test                & Training             & Test                  & Training                & Test                   & Training                & Test                   \\ \hline
Random                     & 81+/-11               & 57+/-9              & 85+/-8               & 66+/-10               & 82+/-9                  & 62+/-7                 & 87+/-12                 & 74+/-23                \\ \hline
CorEx                      & 82+/-19               & 59+/-14             & 83+/-18              & 70+/-11               & 81+/-20                 & 65+/-8                 & 94+/-6                  & 71+/-36                \\ \hline
K-Means                    & 85+/-12               & 53+/-24             & 89+/-10              & 67+/-15               & 79+/-20                 & 56+/-19                & 96+/-3                  & 69+/-38                \\ \hline
Referential {[}28{]}       & 90+/-11               & 59+/-7              & 91+/-9               & 68+/-10               & 90+/-9                  & 67+/-12                & 97+/-4                  & 70+/-35                \\ \hline
\textbf{Proposed}          & \textbf{85+/-11}      & \textbf{68+/-9}     & \textbf{90+/-6}      & \textbf{73+/-13}      & \textbf{82+/-16}        & \textbf{63+/-9}        & \textbf{93+/-6}         & \textbf{89+/-6}        \\ \hline
\end{tabular}
\end{center}
\end{table*}

\subsection{Global Comparison}
\label{section:compa}
\begin{figure}[!t]
\centering
\includegraphics[scale=0.45]{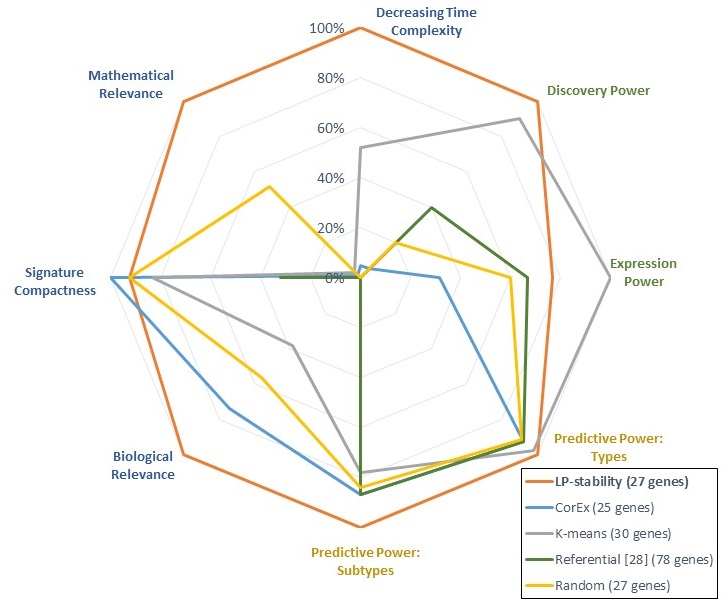}
\caption{\textbf{Comparison of the different signatures.} Blue: criteria based on the gene clustering performance, Green: criteria based on the informativeness of the signature for unsupervised clustering tasks and Gold: criteria based on the relevance of the signature for supervised classification tasks.} 
\label{fig:spider}
\end{figure}

In order to better summarize the different results and provide a fair comparison with random, the state of the art and the referential signatures a spider chart is presented in Fig.~\ref{fig:spider}.  The comparison focuses in  $3$ different criteria: (i) criteria based on the gene clustering performance in blue, (ii) criteria based on the informativeness of the signature for unsupervised clustering tasks in green and (iii) criteria based on the relevance of the signature for supervised classification tasks in gold. Discovery Power is the proportion of tumor types that are relevantly grouped in sample clustering according to related tumor types, the criteria of evaluation are presented in section ~\ref{section:DP}. Expression Power corresponds to the average of the following clustering scores: ARI, NMI, homogeneity, completeness and FMS the results are provided in Supplementary. Predictive Power: Types is the balanced accuracy on test of the tumor types classification task results are provided in Table~\ref{tab:types}. Predictive Power: Subtypes is the average over all tumor types of the balanced accuracy on test for tumor subtypes classification the results are provided in Table~\ref{tab:subtypes}. Biological Relevance is the average ES of the gene clustering method results provided in Table~\ref{tab:bests}. Mathematical Relevance is the average DI score of the gene clustering method results provided in Table~\ref{tab:bests}. Decreasing Time Complexity is the average time taken for the gene clustering, the bigger the area in the chart the faster, results are provided in Table~\ref{tab:bests}.  
Our proposed signature is shown to be largely superior by at least $10\%$ to random and referential signatures in all criteria except compactness. It is also superior to the other signatures designed using our pipeline with other prominent clustering methods. One interesting exception is the Tumor-Specific Expression Power of K-Means-derived signature. The signature defined with K-Means differentiates the types of tumor well as also proved by the Predictive Power: Types but does not perform well on  identifying the subtypes (Predictive Power: Subtypes). This is also due to the lower Discovery Power of K-Means compared to our proposed signature. 

\section{Conclusions}
In this paper, we present a framework for gene clustering definition and comparison, for gene signature selection and evaluation in terms of redundancy, compactness and expression power.  In particular, we present a mathematical and biological evaluation of gene clustering, an extensive sample clustering evaluation using quantitative and field specific clinical, biological metrics, and a supervised approach for its association with tumor types and subtypes characterization. In this framework we have shown the interest of using LP-Stability algorithm, a powerful center-based clustering algorithm, for gene clustering. 
Moreover, the obtained clusters formulate a gene signature 
proving causality and strong associations with tumor phenotypes. These results compete with those reported in the literature by using a large set of different omics data. In addition, our compact signature has been compared and proved to be more expressive than a prominent knowledge-based gene signature~\cite{thorsson2018}. An extensive biological analysis evidenced that the designed signature, leads to sample clusters with high relevance and correlation to cancer-related processes and immune response reporting promising results in tumor types and subtypes classification 
In the future, we aim to extend the proposed method towards discovering stronger gene dependencies through higher-order relations between gene expression data, as well as further evaluation of this biomarker for therapeutic treatment selection in the context of cancer.

\small{
\section{Acknowledgement}
We would like to acknowledge the support of Amazon Web Services grant 
and Pr. Stefano Soatto for fruitful discussions.
We also acknowledge ARC sign'it program and Siric Socrates INCA.
This work was partially supported by the Fondation pour la Recherche M\'edicale (FRM; no. DIC20161236437) and by the ARC sign'it grant ARC: Grant SIGNIT201801286.
We also thank Y. Boursin, M. Azoulay and Gustave Roussy Cancer Campus DTNSI team for providing the infrastructure resources used in this work.
}

\ifCLASSOPTIONcaptionsoff
  \newpage
\fi



%
\bibliographystyle{IEEEtran}
\bibliography{bib} 

\begin{IEEEbiographynophoto}{Enzo Battistella}
holds a MSc and a computer science engineer degree from Telecom Paris. He is currently pursuing a PhD in CentraleSupelec and Gustave Roussy about gene clustering and gene signature designing. He is under the joint supervision of Eric Deutsch and Nikos Paragios.
\end{IEEEbiographynophoto}
\vspace{-1.5cm}
\begin{IEEEbiographynophoto}{Maria Vakalopoulou}
 is an Assistant Professor at MICS laboratory of CentraleSupelec, University Paris-Saclay, Paris, France. Prior to that and during 2017 – 2018, she was a postdoctoral researcher at the same university. She received her PhD, in 2017, from the National Technical University of Athens, Athens, Greece from where she also received her Engineering Diploma degree, graduating with excellence. Her research interests include medical imagery, remote sensing, computer vision, and machine learning. The researcher has published her research in international journals (Lancet Oncology, Radiology, European Radiology) and conferences (MICCAI, ISBI, IGARSS). 

\end{IEEEbiographynophoto}
\vspace{-1.5cm}
\begin{IEEEbiographynophoto}{Roger Sun}
is a resident in radiation oncology at Assistance Public des Hôpitaux de Paris. He is currently doing his PhD under the supervision of Eric Deutsch in the INSERM U1030 molecular radiotherapy laboratory.
\end{IEEEbiographynophoto}

\begin{IEEEbiographynophoto}{Th\'eo Estienne}
holds a MSc and a engineer degree in Applied Mathematics from Centrale Paris (became CentraleSupelec). He is currently pursuing a PhD in collaboration between CentraleSupelec (MICS laboratory) and Gustave Roussy. His topics are medical imaging, deep learning and more specifically the extraction of new bio-markers for brain cancer using deep learning and brain MRI registration based on deep learning.

\end{IEEEbiographynophoto}
\vspace{-1.5cm}
\begin{IEEEbiographynophoto}{Lerousseau Marvin}
is a PhD student in oncology and artificial intelligence.

\end{IEEEbiographynophoto}
\vspace{-1.5cm}
\begin{IEEEbiographynophoto}{Sergey Nikolaev}
 main research's interest has evolved from evolutionary genetics to genetics of cancer, benign tumors and non-tumoral malformative disease.
 In the last 8 years, he has conducted successful genetic research with next generation sequencing on various cancer types, which resulted in important scientific advances published in Nat. Communications, Journal of Pathology, Cancer Research, Nat. Genetics and New England Journal of Medicine. From January 2018 he started his own group in Gustave Roussy Comprehensive Cancer Center located in Villejuif, Paris. In Gustave Roussy his team studies the genetics of Non-Melanoma Skin Cancers and continues to collaborate with clinicians to study the genetics of other rare tumor types and somatic mutations in non-tumoral malformative diseases.


\end{IEEEbiographynophoto}
\vspace{-1.5cm}
\begin{IEEEbiographynophoto}{\'Emilie Alvarez Andres}
holds a MSc in medical physics from the University of Paris-Sud (2017). She is currently pursuing a PhD in collaboration between TheraPanacea and Gustave Roussy about brain MRI-only radiotherapy based on Deep Learning.

\end{IEEEbiographynophoto}
\vspace{-1.5cm}
\begin{IEEEbiographynophoto}{Alexandre Carr\'e}
holds a MSc in medical physics from the University of Paris Saclay. He is currently pursuing a PhD in Gustave Roussy on the analysis of tumor heterogeneity for the personalization of brain tumor treatment plans in radiotherapy.

\end{IEEEbiographynophoto}
\vspace{-1.5cm}
\begin{IEEEbiographynophoto}{St\'ephane Niyoteka}
 is a PhD student at Gustave Roussy about radiomic signature of lymphocyte infiltrate for patients  with locally advanced cervical cancers treated with combined chemoradiation +/- immunotherapy. He holds a MSc in medical physics from the university Paris Saclay.

\end{IEEEbiographynophoto}
\vspace{-1.5cm}
\begin{IEEEbiographynophoto}{Charlotte Robert}
 is an assistant professor at Paris-Saclay University, where she teaches medical and nuclear physics at the Faculties of Medicine and Sciences. After a thesis in physics dedicated to the optimization of semi-conductor SPECT systems, she completed her training with a three-year post-doctoral position at IMNC laboratory (Imaging and Modelization for Neurology and Oncology) in Orsay, France. This post-doc was focused on the use of PET acquisitions for the dose delivery control in hadrontherapy. Since 2019, she is coordinating research activities dedicated to artificial intelligence and multi-modal imaging for personalized radiotherapy in the radiation therapy department of Gustave Roussy Hospital.
\end{IEEEbiographynophoto}
\vspace{-1.5cm}
\begin{IEEEbiographynophoto}{Nikos Paragios}
received his B.Sc. and M.Sc.degrees from the University of Crete, Gallos, Greece,in 1994 and 1996, respectively, his Ph.D. degree from I.N.R.I.A./University of Nice/Sophia Antipolis, Nice,France, in 2000 and his HDR (Habilitation \`a Diriger des Recherches) degree from the same university in 2005. He is a Professor of applied mathematics at CentraleSupelec/ Universit Paris-Saclay Orsay, France. He has published more than 200 papers in the areas of computer vision, biomedical imaging, and machine learning. Prof. Paragios is the Editor-in-Chief of the Computer Vision and Image Understanding journal and serves on the editorial board of the Medical Image Analysis and SIAM Journal on Imaging Sciences. He is a Senior Fellow of the Institut Universitaire de France and a Member of the Scientific Council of Safran conglomerate.
\end{IEEEbiographynophoto}
\vspace{-1.5cm}
\begin{IEEEbiographynophoto}{\'Eric Deutsch}
is an oncologist-radiotherapist, head of the Radiotherapy Department of Gustave Roussy and director of the joint research unit "Molecular Radiotherapy" (Inserm - University Paris-XI). 

His research has been dedicated to combination of novel anticancer drugs either used alone or in combination to radiotherapy. Prof. Deutsch’s interest focuses on cell death mechanisms, inflammation and immunotherapy, on the clinical side, he is versed into the field of translational research and early clinical trials. He has investigated several first in human novel drugs-radiotherapy combinations such as mTOR inhibitors, antiviral agents, immune modifiers and nanoparticles. Research axes: Early clinical trials and phase I, nanoparticles, oligometastasis,  , radiation biology, HPVs related tumors, combination with new drugs and Radiotherapy.

In addition, his clinical research focuses on developing new irradiation processes to better target and eliminate tumors. By using technological innovations such as  radiomics, computational imaging, guidance by medical imaging or novel anti-cancer agents Prof. Deutsch works with his medical team on increasingly personalized protocols that can adapt to very specific cancers. Together with members of the research unit he leads, Prof. Deutsch studies the biological effects of irradiation. The objectives of this research are first to identify the biological characteristics of tumors, "tumor markers", to predict their sensitivity or resistance to radiotherapy.

\end{IEEEbiographynophoto}




\end{document}


\title{Supplementary Materials}
\maketitle
\section{Formal Definition of Proximity Measures}

We detail here the difference distance or similarity metrics used in this article.
\begin{itemize}
\item \textbf{Euclidean Distance:} Euclidean distance is one of the most commonly used metrics, measuring the dissimilarity between vectors. 

$euclidean(x^p,x^q)=\lVert x^p - x^q \rVert_2$

\item \textbf{Cosine Distance:} Cosine distance is also a distance that is very commonly used in the literature. 
It can be defined as follows:

$cosine(x^p,x^q) =1 - \dfrac{x^p\cdot x^q}{\lVert x^p \rVert_2 \lVert x^q \rVert_2}$

\item  \textbf{Pearson's correlation:} Pearson's correlation is based on the covariance. It assesses very well the linear correlations between the variables.

$Pearson(x^p,x^q) = \dfrac{cov(x^p,x^q)}{\sigma(x^p) \sigma(x^q)}$.

\item \textbf{Spearman's  rank correlation:} Spearman's correlation is a non-parametric measure of rank correlation and it assesses to what extent the relation between two variables can be represented by a monotonic function.
It is calculated by: 

$Spearman(u,v) = \dfrac{cov(rg(x^p),rg(x^q))}{\sigma(rg(x^p)) \sigma(rg(x^q))}$.

\item \textbf{Kendall's rank correlation:} Similarly to Spearman's correlation, Kendall's correlation is a measure of rank correlation considering the similitude of the ranking order of the observations for the two compared objects. It assesses the best non-linear dependencies. 
It is calculated by:   

$Kendall(x^p,x^q)=2\dfrac{N_C - N_D}{n(n-1)}$ where $N_C$ is the number of concordant pairs and $N_D$ the number of discordant pairs. Pairs of observations $(x^p_u, x^p_v)$ and $(x^q_u,x^q_v)$ are considered concordant if their ranks agree i.e. $x^p_u > x^p_v \Leftrightarrow x^q_u > x^q_v$ they are said discordant if $x^p_u > x^p_v \Leftrightarrow x^q_u < x^q_v$.

\item \textbf{Kullback-Leibler divergence:} Kullback-Leibler divergence measures how different two probability distributions are. It represents the expectation that two distributions present a similar behavior.
It is computed by: 

$KL(x^p,x^q)=-\sum_u P(x^p_u) \log \dfrac{P(x^q_u)}{P(x^p_u)}$. We made this measure symmetric by considering 

$KL(x^p,x^q) + KL(x^q,x^p)$.
 
\end{itemize}
\section{Formal Definition of Clustering Evaluation Metrics}
A number of different metrics were used to evaluate the different clustering algorithms for the sample clustering. In this section, we present formally the different metrics.
\begin{enumerate}
    \item \textbf{Adjusted Rand Index (ARI):} It is a similarity measure between a clustering $C'$ and a ground truth $C$. RI corresponds to the proportion of pairs of elements that are in different clusters in both $C$ and $C'$ called $a$ or in the same cluster in both $C$ and $C'$ called $b$. 
    \begin{equation}
        RI = \dfrac{a+b}{\dbinom{n}{2}}
    \end{equation}
    RI is then corrected for chance by taking into account its expected value $E(RI)$:
    \begin{equation}
        ARI = \dfrac{RI - E(RI)}{max(RI) - E(RI)}
    \end{equation}
    
    \item \textbf{Normalized Mutual Information (NMI):} It is a normalized mutual information metric between a clustering $C'$ and the ground truth class $C$.  \begin{equation}
        NMI = \dfrac{MI(C,C')}{mean(H(C),H(C'))}
    \end{equation}
    Mutual Information (MI) and Shannon Entropy (H) are defined in Section III-A of the manuscript.
    \item \textbf{Homogeneity:} Considering a clustering $C'$ and the ground truth $C$. It values clusters of $C$ containing elements all belonging to a same cluster in $C'$.
    \begin{equation}
        homogeneity = 1 - \dfrac{H(C|C')}{H(C)}
    \end{equation}
    \item \textbf{Completeness:} Considering a clustering $C'$ and the ground truth $C$, completeness is a complement of homogeneity as it values clusters of $C$ having all their elements belonging to the same cluster in $C'$.
    \begin{equation}
        completeness = 1 - \dfrac{H(C'|C)}{H(C')}
    \end{equation}
    
    \item \textbf{Fowlkes-Mallow Score (FMS):} Considering a clustering $C'$ and the ground truth clustering $C$. This metric corresponds to the geometric mean of the pairwise precision and recall.
    \begin{equation}
        FMS = \dfrac{TP}{\sqrt{(TP+FP)(TP+FN)}}
    \end{equation}
    the $TP$, $FP$ and $FN$ indicate the numbers of true positives, false positives and false negatives respectively.
\end{enumerate}

\section{Computational Complexity \& Running Times For Gene Clustering}
\label{section:time}
The computation time is an important parameter playing a significant role for the selection of a clustering algorithm. For each algorithm, the approximate average time needed for the clustering is presented in Table~II of the main manuscript. The different computational times have been computed using Intel(R) Xeon(R) CPU E5-4650 v2 @ 2.40GHz cores. In general, the computational time increases with the cluster number for all the clustering methods. However, for the reported clusters of Table~II, LP-Stability remains one of the fastest with a computational time approximately equal to 1.5h.  K-Means needs approximately twice this time due to the several iterations (in our case $100$) performed in order to account for different initialization conditions. CorEx is by far the most computationally expensive, requiring more than 5 days for the clustering, making this algorithm not efficient for data with this high dimensionality.

\section{Gene Signature Composition}
We detail here the $27$ genes of our proposed signature and their main functions. We also provide a brief summary of the analysis obtained using GTEx Portal on July 2020 (www.gtexportal.org).
\begin{itemize}
    \item HSFX1: DNA binding transcription, GTEx: Overexpressed in brain cerebellum and cerebellar hemisphere and ovary tissues
    \item C3P1: endopeptidase inhibitor activity, GTEx: Highly overexpressed in liver tissues
    \item CCDC30: Coiled-Coil Domain, GTEx: Slightly overexpressed in all brain tissues
    \item CNRIP1: cannabinoid receptor, GTEx: Particularly expressed in many tissues and in particular all brain tissues
    \item CD53: regulation of cell development GTEx: Highly expressed in blood and lymphocytes
    \item SPRR4: UV-induced cornification, GTEx: more expressed in sun exposed tissues particularly skin
    \item RIF1: DNA repair, GTEx: expressed in many tissues including heart, blood lymphocytes and brain
    \item COL1A2: collagen making, GTEx: highly overexpressed in cultured fibroblasts
    \item ZNF767: gene expression, GTEx: highly expressed in several tissues including uterus, vagina, ovary, brain cerebellum and cerebellar hemisphere
    \item CD3E: antigen recognition (linked to immunodeficiency), GTEx: more expressed in whole blood tissues
    \item MATR3: nucleic acid binding and nucleotide binding, GTEx: highly expressed in several tissues including uterus, vagina, ovary and brain
    \item NCAPH: Cell Cycle, Mitotic and Mitotic Prometaphase, GTEx: highly expressed in EBV-transformed lymphocytes and on a smaller extend in cultured fibroblasts
    \item ASH1L: transcriptional activators,  GTEx: expressed in many tissues including heart, blood lymphocytes,  uterus, vagina, ovary and brain
    \item ANKRD30A: DNA-binding transcription factor activity (related to breast cancer), GTEx: more expressed in breast mammary tissues
    \item GNA15: among its related pathways are CREB Pathway and Integration of energy metabolism, GTEx: especially overexpressed in oesophagus mucosa
    \item GADD45GIP1: Among its related pathways are Mitochondrial translation and Organelle biogenesis and maintenance, GTEx: expressed in many tissues slightly overexpressed in cultured fibroblasts
    \item CD302: cell adhesion and migration, GTEx: especially overexpressed in lung and liver tissues
    \item SFTA3: Among its related pathways are Surfactant metabolism and Diseases of metabolism, GTEx: Overexpressed in Lung and thyroid
    \item C1orf159: Protein Coding gene, GTEx: especially overexpressed in testis
    \item RPS8: Among its related pathways are Viral mRNA Translation and Activation of the mRNA upon binding of the cap-binding complex and eIFs, and subsequent binding to 43S, GTEx: especially overexpressed in ovary tissues
    \item ZEB2: Among its related pathways are MicroRNAs in cancer and TGF-beta Receptor Signaling, GTEx: especially overexpressed in spinal cord (c-1) and tibial nerve
    \item GSX1: sequence-specific DNA binding and proximal promoter DNA-binding transcription activator activity, RNA polymerase II-specific, GTEx: especially expressed in hypothalamus
    \item ADNP: Vasoactive intestinal peptide is a neuroprotective factor that has a stimulatory effect on the growth of some tumor cells and an inhibitory effect on other, GTEx: overexpressed in quantity of tissues, especially so in EBV-transformed lymphocytes, testis, ovary and uterus
    \item CLIP3: plays a role in T cell apoptosis by facilitating the association of tubulin and the lipid raft ganglioside GD3, GTEx: expressed in many tissues slightly overexpressed in EBV-transformed lymphocytes
    \item YEATS2: Among its related pathways are Chromatin organization, GTEx: expressed in many tissues overexpressed in EBV-transformed lymphocytes
    \item ACBD4: Among its related pathways are Metabolism and Peroxisomal lipid metabolism, GTEx: expressed in many tissues overexpressed in liver, thyroid, uterus and vagina
    \item SNRPG: Among its related pathways are mRNA Splicing - Minor Pathway and Transport of the SLBP independent Mature mRNA, GTEx: expressed in many tissues strongly overexpressed in EBV-transformed lymphocytes and cultured fibroblasts
\end{itemize}

\section{Samples Clustering Comparison}
In Fig.~1, we present the sample clustering performed using the CorEx signature having maximal ES and DI. This clustering due to the very low number of clusters failed to provide a good sample clustering, giving quite intermixed clusters. This is the reason that for further evaluation of the CorEx algorithm we used the gene signature provided by 25 genes (Fig.~5 and of the main manuscript). Moreover in Fig.~2 we present the influence of two different distance metrics in the sample clustering of the same gene signature (LP-Stability with 27 genes). In particular, Spearman’s correlation tends to better separate the different tumors into different clusters, while the Kendall’s seems to generate clusters that groups tumor-related sample.

\begin{figure}[!t]
\centering
\includegraphics[scale=0.48]{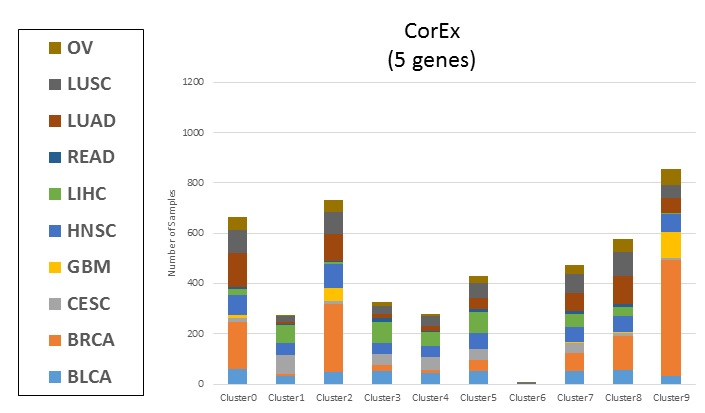}
\caption{\textbf{Gene Signature Assessment for the CorEx algorithm.} The graph depicts the distribution of the different tumor types in $10$ different clusters using the best signature produced by CorEx algorithm ($5$ genes). From the graph one can observe that the different tumor types are quite intermixed across the different clusters without any association between them.  }
\label{fig:signature_fail}
\end{figure}


\begin{figure*}[!t]
\centering
\includegraphics[scale=0.40]{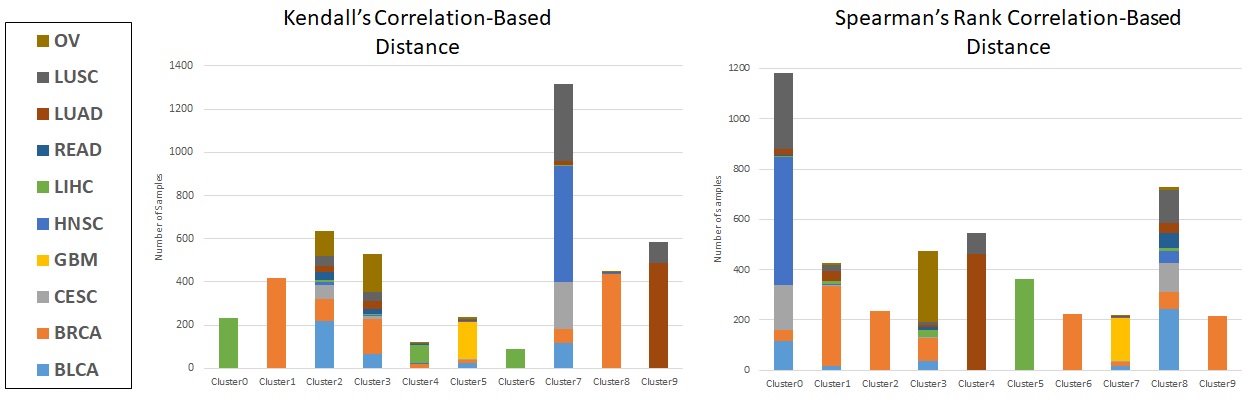}
\caption{\textbf{Gene Assessment performed with LP-Stability clustering and Kendall's correlation-based distance.}
  The plot presents the distribution of tumors using the signature produced by LP-Stability and Kendall's correlation-based distance (right) in comparison to the Spearman's one (left). This assessment is performed in order to compare the influence of the distance in the clustering. We can observe that there are some similar clusters between the two distances such as the well defined clusters of GBM and LIHC together with LUAD/LUSC cluster (Cluster $5$), a squamous cluster (Cluster $7$) and some well defined BRCA clusters (Clusters $1$ and $8$). Thus, regarding sample clustering, it appears that the good characterization of monotonic relations offered by Spearman's Rank correlation-based distance is better suited than the more general characterization of the Kendall's one.}
\label{fig:sample_kendall}
\end{figure*}

In Table~I, we present a more detailed comparison of the  distribution of the tumor types for these LP-Stability and K-Means signatures. Both signatures generate clusters that successfully associate lung tumors such as LUSC and LUAD (clusters 3 \& 4 respectively), squamous tumors mainly composed of  BLCA, CESC, LUSC and HNSC types (clusters 0 \& 8 and 1 \& 8 respectively) and smoking related tumors mainly containing CESC, HNSC, READ, LUSC and LUAD (clusters 7 \& 8 respectively). Concerning BRCA,  K-Means clusters it into two different groups, one that consists mainly with BRCA samples, while the second one consists of a minority of BRCA samples grouped together with the GBM which types are not really related.
On the other hand, LP-Stability clusters BRCA in several small unblended clusters that express the various molecular types of BRCA, and groups the remaining BRCA with the OV type which is directly related (cluster 3). These results are very promising as they are in accordance with other recent omic studies. In particular in [2] the authors used a large set of different omics data to define a clustering reporting pan-squamous clusters (LUSC, HNSC, CESC, BLCA), but also pan-gynecology clusters (BRCA, OV) and pan-lung clusters (LUAD, LUSC). They also reported the separation of BRCA into several clusters linked to basal, luminal, Chr 8q amp or HER2-amp subtypes.
Moreover, both algorithms provided a good, almost perfect separation of the LIHC and GBM samples into well defined clusters. This separation indicates that these specific tumor types are very different from the rest or even that at least one gene included in the produced signatures is differently expressed compared to the rest of the samples.

\begin{table*}[!t]
\label{tab:samples_compa}
\caption{Discovery Power: A complete comparison for the distribution of the tumor types (above $10\%$) from the best performing algorithms. LP-Stability with $27$ genes using Kendall's correlation-based distance and K-Means with $30$ genes using Euclidean distance. The last column indicates the algorithm that provided the best distribution for the specific tumor type. It highlights the superiority of the LP-Stability signature.
}
\centering
\begin{tabular}{|c|c|c|c|}
\hline
Tumor Types & \textbf{LP-Stability} ($27$ genes) & \textbf{K-Means} (30 genes) & Best\\ \hline  \hline
BLCA & 
\begin{tabular}{@{}c@{}}
$57\%$ BLCA $\Rightarrow$ $33\%$  cluster 8 \\ 
$26\%$ BLCA $\Rightarrow$ $10\%$  cluster 0 \\  
$< 10\%$ BLCA $\Rightarrow$ clusters 1, 3, 7
\end{tabular} & 
\begin{tabular}{@{}c@{}}
$54\%$ BLCA  $\Rightarrow$ $59\%$  cluster 7 \\
 $18\%$ BLCA $\Rightarrow$ $22\%$ cluster 1 \\ 
 $14\%$ BLCA $\Rightarrow$ $7\%$ cluster 8\\
 $< 10\%$ BLCA $\Rightarrow$ cluster 2, 4, 9\\
 \end{tabular}  
 &  $\sim$\\
\hline
BRCA & 
\begin{tabular}{@{}c@{}}
$26\%$ BRCA  $\Rightarrow$ $75\%$  cluster 1 \\
$20\%$ BRCA  $\Rightarrow$ $100\%$  cluster 2 \\
$19\%$ BRCA $\Rightarrow$ $100\%$ cluster 6 \\ 
$18\%$ BRCA $\Rightarrow$ $100\%$ cluster 9\\
$10\%$ BRCA $\Rightarrow$ $20\%$ cluster 3\\
\textbf{Homogeneous Clusters or with related types}
\end{tabular} & 
\begin{tabular}{@{}c@{}} 
$55\%$ BRCA $\Rightarrow$ $98\%$ cluster 0\\ 
$27\%$ BRCA $\Rightarrow$ $20\%$ cluster 4\\
$< 10\%$ BRCA $\Rightarrow$ clusters 1, 2, 7\\
Clusters unrelated to GBM type
\end{tabular} &
 LP-Stability\\
\hline
CESC &
\begin{tabular}{@{}c@{}} 
$58\%$ CESC $\Rightarrow$ $15\%$ cluster 0\\ 
$38\%$ CESC $\Rightarrow$ $16\%$ cluster 8\\
\textbf{Squamous related clusters}
\end{tabular} & 
\begin{tabular}{@{}c@{}} 
$54\%$ CESC $\Rightarrow$ $15\%$ cluster 8\\ 
$25\%$ CESC $\Rightarrow$ $16\%$ cluster 1\\
$16\%$ CESC $\Rightarrow$ $16\%$ cluster 7\\
\textbf{Squamous mixed with non squamous}
\end{tabular} &
 LP-Stability\\
\hline
GBM  & 
$100\%$ GBM $\Rightarrow$ $79\%$ cluster 7 & 
\begin{tabular}{@{}c@{}}
$98\%$ GBM $\Rightarrow$ $57\%$ cluster 2 \\
Mixed with unrelated BRCA types
\end{tabular} 
& LP-Stability\\
\hline
HNSC & 
\begin{tabular}{@{}c@{}} 
$89\%$ HNSC $\Rightarrow$ $43\%$ cluster 0\\ 
$10\%$ HNSC $\Rightarrow$ $7\%$ cluster 8\\
\textbf{Squamous related clusters}
\end{tabular} & 
\begin{tabular}{@{}c@{}} 
$86\%$ HNSC $\Rightarrow$ $62\%$ cluster 8\\ 
$11\%$ HNSC $\Rightarrow$ $18\%$ cluster 1\\
\textbf{Squamous related clusters}
\end{tabular} &
$\sim$\\
\hline
LIHC & $90\%$ LIHC $\Rightarrow$ $100\%$ cluster 5 &
$98\%$ LIHC $\Rightarrow$ $98\%$ cluster 5 &
$\sim$\\
\hline
READ &
\begin{tabular}{@{}c@{}} 
$82\%$ READ $\Rightarrow$ $9\%$  cluster 8
\\ \textbf{Smoking related} 
\end{tabular} & 
\begin{tabular}{@{}c@{}}
$55\%$ READ $\Rightarrow$ $10\%$  cluster 7 \\
$32\%$ READ $\Rightarrow$ $5\%$  cluster 4 
\\ \textbf{Smoking related}
\end{tabular} & 
$\sim$\\
\hline
LUAD & 
\begin{tabular}{@{}c@{}}
$80\%$ LUAD $\Rightarrow$ $85\%$ cluster 4 \\  
\textbf{Lung cluster}
\end{tabular} & 
\begin{tabular}{@{}c@{}}
$93\%$ LUAD $\Rightarrow$ $83\%$ cluster 3\\
\textbf{Lung cluster}
\end{tabular}  & 
$\sim$\\
\hline
LUSC & 
\begin{tabular}{@{}c@{}} 
$54\%$ LUSC $\Rightarrow$ $25\%$ cluster 0\\ 
$23\%$ LUSC $\Rightarrow$ $18\%$ cluster 8 \\
$15\%$ LUSC $\Rightarrow$ $15\%$ cluster 4\\
\textbf{Squamous and lung clusters}
\end{tabular}  & 
\begin{tabular}{@{}c@{}} 
$53\%$ LUSC $\Rightarrow$ $97\%$ cluster 6\\ 
$20\%$ LUSC $\Rightarrow$ $17\%$ cluster 3 \\
$11\%$ LUSC $\Rightarrow$ $21\%$ cluster 1\\
\textbf{Squamous and lung clusters}
\end{tabular} & 
K-Means\\
\hline
OV   &
\begin{tabular}{@{}c@{}} 
$92\%$ OV  $\Rightarrow$ $60\%$ cluster 3\\ 
$< 5\%$ OV $\Rightarrow$ clusters 1, 8\\
\textbf{Cluster with related BRCA}
\end{tabular}  & 
\begin{tabular}{@{}c@{}}
$71\%$ OV  $\Rightarrow$ $86\%$ cluster 9 \\ 
$15\%$ OV  $\Rightarrow$ $10\%$ cluster 4 \\
$10\%$ OV  $\Rightarrow$ $7\%$ cluster 7 \\
$< 10\%$ OV $\Rightarrow$ clusters 0,2\\
\textbf{Mixed clusters}
\end{tabular} & 
LP-Stability\\ \hline
\end{tabular}
\end{table*}
\section{Gene Screening Analysis for K-Means algorithm ($30$ genes).}
In this section we present a detailed gene screening analysis for the different clusters produced by the K-Means signature and presented in Fig.~5 on the main manuscript. 
We noticed that cluster $0$ which is a well defined cluster containing mainly BRCA samples, presents enrichment in diverse biological processes as regulation of transcription by RNA polymerase II, regulation of nucleobase-containing compound metabolic process or regulation of gene expression. The most significant gene C10orf32 has not been identified as a gene related to cancer. However, it is more related to the lysosomes movement process.

Cluster $1$ seems to be very intermixed with different biological processes being enriched. 
For the BRCA samples different skin related pathways especially keratin are enriched, which are important for several types of cancers. The most significant gene for BRCA samples PKP1 is related to molecular recruitment. However, different processes for other tumor types are also enriched. In particular, for the LUSC samples do not present any significant gene with high enough score. The one with the highest score is ANKRD13B which is related to membrane binding processes. 
HNSC samples are enriched in general RNA metabolic processes and DNA-binding. Their most significant gene, CADPS2, is involved in calcium binding especially important in autism. BLCA samples have no genes with scores above the considered threshold. The most significant gene is FOXC1 which is involved in DNA-binding and has been shown of utmost interest in several type of cancers. CESC type do not have any significant gene with TRIM8 being the one with the highest score. This gene seems to be related to Interferon gamma signaling and Innate Immune System. Its regulation has been shown to be altered in some cancers. After this analysis, it appears that this cluster contains rather heterogeneous samples without common biological processes even if several are linked to cancer. Besides, the biological relevance of the cluster is not very clear as we can observe very few significant genes per tumor type.

Cluster $2$ groups GBM and BRCA samples. It presents for the BRCA ones an enrichment in voltage-gated calcium channel activity only. This biological pathway has been identified as a new target for BRCA in~[3]. The most enriched gene is CACNB2 which is an antigene involved in voltage-gated calcium channel. A study for the GBM samples cannot be performed as all the GBM samples are in this cluster.

Regarding cluster $3$, LUSC samples present enrichment in cilium activity and surfactant homeostasis. Their most significant gene ARRB1 programs a desensitization to stimuli. It seems to be of interest for the chemosensitivity of lung cancer. For the LUAD samples in this cluster, the only significant gene NKX2-1 is a thyroid-specific gene also involved in morphogenesis. It has been found to be a prognostic marker in early stage non-small lung cancers.

Cluster $4$ consists mainly of BRCA samples which are enriched in processes of immune response. However, the most significant gene ACTR3 code for a complex essential for cell shape and motility which is not related to immune response. Cluster $4$ seems quite heterogeneous concerning the processes and the significant genes. In particular, the HNSC samples are related to extra-cellular organization. Their most significant gene KLF17 is related to DNA-binding transcription that is involved in epithelial-mesenchymal transition and metastasis in breast cancer. READ samples do not have significant genes, the one with the highest score is the GRM2 which is particularly involved in neurotransmission and central nervous diseases. For the OV samples the only significant gene is the SPHK1 which regulates cell proliferation and cell survival. It has been linked to ovarian cancer in~[1].
LUAD samples present few significant genes which do not enrich any biological process. The most significant gene, UCA1, plays a role in cell proliferation and has been proven to be of interest in bladder cancer.

Cluster $5$ groups the entire LIHC tumor type. Thus, a gene screening analysis is so not possible.

For the cluster $6$, LUSC samples have very few significant genes. The enriched processes that are associated with are related to tissue development, Estrogen signaling and mammary gland morphogenesis. Its most significant gene is FRRS1 which is related to ferric-chelate reductase activity. Thus, this homogeneous cluster does not seem to contain a biological meaningful subset of LUSC samples.

\begin{table*}[!t]
\caption{\textbf{Expression Power} of the sample clustering using as features respectively our proposed signature, a referential signature from literature~[28] and average performance using $10$ sets of randomly-selected genes of same size as the proposed signature. We observe that the two best performing signatures are the ones produced with our pipeline. The first using K-Means clustering the second, our proposed signature, using LP-Stability.}
\label{tab:EP}
\centering
\begin{tabular}{|c|c|c|c|c|c|c|}
\hline
Signature         & ARI (\%)          & NMI (\%)           & Homogeneity (\%)  & Completeness (\%) & FMS (\%) & \begin{tabular}[c]{@{}l@{}}Expression \\ Power (\%) \end{tabular} \\  \hline
Random            & 29+/-5    & 37+/-4    & 37+/-4    & 37+/-4    & 39+/-4     & 36             \\ \hline
CorEx             & 12          & 20          & 21          & 20          & 23           & 19             \\ \hline
K-Means           & 52          & 63          & 65          & 62          & 58           & 60             \\ \hline
Referential       & 34          & 41          & 42          & 40          & 42           & 40             \\ \hline
\textbf{Proposed} & \textbf{33} & \textbf{52} & \textbf{52} & \textbf{53} & \textbf{43}  & \textbf{46}   
    \\ \hline
\end{tabular}
\end{table*}

Cluster $7$ groups BLCA, BRCA, CESC, LUSC and READ samples.
OV samples present very few significant genes without enriched biological processes.
BLCA samples present significant genes related to transcription and the most significant gene, C17orf28, is related to several cancers. BRCA samples have very few significant genes and are weakly enriched in mitosis processes as there are only two enriched processes. The most significant gene FSD1 is related to coiled-coil region. CESC samples are enriched in cilium organization, cell projection assembly and the most significant gene is EPCAM which is related to gastrointestinal carcinoma and is a target of immunotherapy. So, it does not present links with CESC or other carcinomas of the cluster. It seems that these CESC samples would have been more suitable for cluster $3$ since LUSC samples of this cluster are strongly enriched in the same pathways. Finally, READ samples do not present significant enough genes. However, the most significant one is EFNB2 which is involved in several development processes and in particular in the nervous system and in erythropoiesis. This gene has also been found of interest in tumor growth.

For cluster $8$, LUSC samples have numerous significant genes enriched in epidermis related processes and skin development pathways and in particular the most significant gene KRT14 is related to these processes. This might be related to a subset of non-small cell lung cancers characterized by Epidermal growth factor receptor (EGFR) mutations. Similarly, HNSC samples of cluster $8$ are also linked to keratin, epidermis and skin development processes which also characterize a subtype of HNSC. The most significant gene lad1 is related to structural molecule activity and codes for a protein involved in the basement membrane zone.
We found the same pathways for CESC samples whose most significant gene is KRT5 and BLCA samples with KRT6A.

Cluster $9$ mainly contains OV samples which do not present gene significant enough. However, their most significant gene CLU has been identified as a potential cancer target in~[4].


\section{Expression Power}
We report here the detailed results of section VII-B-4 of the main manuscript. In particular in Table~II, we report the performance of the ARI, NMI, Homogeneity, Completeness, FMS and overall Expression Power (average of the different metrics). One could observe that the best performance is reported by K-Means and our proposed signature produced with LP-Stability. K-Means superiority in different metrics is indicated due to the very good performance on the BRCA samples, the most represented type in our dataset. Besides, those scores does not take into account the biological meaning of the clusters however they offer a general indication of the well-definition of the sample clustering.

\section{Implementation Details For Tumor/ Subtumor Classification}
 Regarding the supervised categorization of tumor types and subtypes classes, the evaluated algorithms were: Nearest Neighbor, \{Linear, Sigmoid, Radial Basis Function (RBF), Polynomial Kernel\} Support Vector Machines (SVM), Gaussian Process, Decision Trees, Random Forests, AdaBoost, XGBoosting, Gaussian Naive Bayes, Bernoulli Naive Bayes, Multi-Layer Perceptron (MLP) \& Quadratic Discriminant Analysis. 
We selected the top classifiers regarding balanced accuracy ensuring both good performance and good generalisation. In particular, for tumor types classification the selection criteria include \textit{(i)} high balanced accuracy (equal or above $80\%$) on the validation set and \textit{(ii)} small difference (smaller than $20\%$) on the balanced accuracy metric between training and validation. While for tumor subtypes classification, we selected the top $5$ classifiers regarding balanced accuracy presenting a small difference (smaller than $20\%$) on the balanced accuracy metric between training and validation. For our experiments on tumor types classification with our proposed signature, the classifiers that fulfill these criteria were the: \{Linear, Polynomial, RBF Kernels\} SVM, Gaussian Process, Random forest, MLP, XGBoosting. For the sake of conciseness, we do not detail the selected classifiers for other experiments and other signatures. Those top classifiers' good performance were leveraged through a majority voting scheme.

To deal with the problem of the unbalanced dataset, each class received a weight inversely proportional to its size. Concerning the different hyperparameters of the best performing classifiers, SVM was granted a regularization parameter of $10$ and polynomial kernel function of degree $4$ for the Polynomial method. In addition, the RBF SVM was granted a kernel coefficient of $3$. The Gaussian Process was granted a RBF kernel and the multi class predictions were achieved through one versus rest scheme. The Random Forest classifier was composed of $100$ Decision Trees of maximum depth $4$. The MLP classifier was used with a LBFGS optimizer, a ReLU activation, $3000$ maximum iterations, a batch size of $200$, learning rate was updated thanks to an inverse scaling exponent of power $t$ with $t$ denoting the current step and early stopping method was used as the termination criteria. XGBoosting was used with $n_{classes}$ regression trees at each boosting stage, a deviance loss, a learning rate of $0.5$ and $40$ boosting stages, when looking for the best split, $\sqrt{n_{features}}$ features were considered.

\section{Full Tumor Types Classification performance}
In Table~III we present the training/ validation results of the different classifiers using the LP-Stability and our proposed signature. Moreover, in Table~IV, we present the results for the training/ test tumor classification results for our proposed signature. The table reports the performance of the selected algorithms together with the voting (ensemble) classifier.

\begin{table*}[!h]
\caption{\textbf{Predictive Power: Tumor Types, Proposed Signature} Training-Validation tumor types classification performance using the proposed signature ($27$ genes). Voting Classifier is composed of classifiers having reached a balanced accuracy above $80\%$ on validation.}
\centering
\begin{tabular}{|l|c|c|c|c|c|c|c|c|}
\hline
\multirow{2}{*}{Classifier}        & \multicolumn{2}{l|}{Balanced   Accuracy (\%)} & \multicolumn{2}{l|}{Weighted   Precision (\%)} & \multicolumn{2}{l|}{Weighted   Sensitivity (\%)} & \multicolumn{2}{l|}{Weighted   Specificity (\%)} \\ \cline{2-9} 
                                   & Training            & Validation         & Training            & Validation          & Training             & Validation           & Training             & Validation           \\ \hline
Nearest Neighbors          & 88               & 79               & 92                & 86               & 92                 & 85                & 97                 & 95                \\  \hline
Linear SVM                 & 91               & 88               & 91                & 90               & 89                 & 89                & 99                 & 99                \\ \hline
poly SVM                   & 98               & 91               & 97                & 92               & 96                 & 92                & 100                & 99                \\ \hline
sigmoid SVM                & 55               & 50               & 70                & 67               & 50                 & 49                & 94                 & 94                \\ \hline
RBF SVM                    & 98               & 89               & 96                & 91               & 96                 & 90                & 100                & 98                \\ \hline
Gaussian Process           & 96               & 90               & 97                & 94               & 97                 & 93                & 99                 & 98                \\ \hline
Decision Tree              & 68               & 66               & 85                & 38               & 47                 & 45                & 94                 & 94                \\ \hline
Random Forest              & 93               & 89               & 94                & 92               & 92                 & 90                & 99                 & 99                \\ \hline
MLP                        & 100              & 87               & 100               & 92               & 100                & 92                & 100                & 98                \\ \hline
AdaBoost                   & 72               & 64               & 81                & 75               & 74                 & 70                & 98                 & 95                \\ \hline
Gaussian Naive Bayes       & 32               & 32               & 69                & 61               & 58                 & 58                & 69                 & 69                \\ \hline
Bernouilli Naive Bayes     & 59               & 59               & 75                & 71               & 74                 & 75                & 90                 & 91                \\ \hline
QDA                        & 71               & 67               & 87                & 82               & 78                 & 76                & 98                 & 98                \\ \hline
XGBoosting                 & 100              & 88               & 100               & 93               & 100                & 92                & 100                & 98                \\ \hline
\textbf{Voting Classifier} & \textbf{99}      & \textbf{92}      & \textbf{98}       & \textbf{94}      & \textbf{98}        & \textbf{94}       & \textbf{100}       & \textbf{99} \\ \hline
\end{tabular}
\end{table*}

\begin{table*}[!h]
\caption{\textbf{Predictive Power: Tumor Types, Proposed Signature} Training-Test tumor types classification performance using the proposed signature ($27$ genes) after retraining on entire Training-Validation set}
\centering
\begin{tabular}{|l|c|c|c|c|c|c|c|c|}
\hline
\multirow{2}{*}{Classifier}        & \multicolumn{2}{l|}{Balanced   Accuracy (\%)} & \multicolumn{2}{l|}{Weighted   Precision (\%)} & \multicolumn{2}{l|}{Weighted   Sensitivity (\%)} & \multicolumn{2}{l|}{Weighted   Specificity (\%)} \\ \cline{2-9} 
                                   & Training            & Test               & Training            & Test                & Training             & Test                 & Training             & Test                 \\ \hline
Linear SVM                 & 91               & 89               & 91                & 90               & 89                 & 88                & 99                 & 98                \\\hline
poly SVM                   & 98               & 87               & 97                & 89               & 97                 & 88                & 100                & 98                \\\hline
RBF SVM                    & 98               & 88               & 97                & 90               & 97                 & 88                & 100                & 99                \\\hline
Gaussian Process           & 95               & 88               & 97                & 92               & 97                 & 92                & 99                 & 98                \\\hline
Random Forest              & 92               & 90               & 93                & 92               & 91                 & 90                & 99                 & 99                \\\hline
MLP                        & 100              & 87               & 100               & 90               & 100                & 89                & 100                & 98                \\\hline
XGBoosting                 & 100              & 91               & 100               & 94               & 100                & 94                & 100                & 98                \\\hline
\textbf{Voting Classifier} & \textbf{99}      & \textbf{92}      & \textbf{99}       & \textbf{94}      & \textbf{98}        & \textbf{93}       & \textbf{100}       & \textbf{99}      \\ \hline
\end{tabular}
\end{table*}

Tables~V and VI we summarise the performances for the signature presented in~[28]. Using the referential algorithm~[28] only three classifiers were selected and used for the tumor classification, reporting also lower performance.

\begin{table*}[!h]
\caption{\textbf{Predictive Power: Tumor Types, Referential Signature~[28]} Training-Validation tumor types classification performance using the referential signature ($78$ genes). Voting Classifier is composed of classifiers having reached a balanced accuracy above $80\%$ on validation and presenting a difference of balanced accuracy between training and validation below $20\%$.}
\centering
\begin{tabular}{|l|c|c|c|c|c|c|c|c|}
\hline
\multicolumn{1}{|c|}{\multirow{2}{*}{Classifier}} & \multicolumn{2}{c|}{Balanced   Accuracy (\%)} & \multicolumn{2}{c|}{Weighted   Precision (\%)} & \multicolumn{2}{c|}{Weighted   Sensitivity (\%)} & \multicolumn{2}{c|}{Weighted   Specificity (\%)} \\ \cline{2-9} 
\multicolumn{1}{|c|}{}                            & Training            & Validation         & Training            & Validation          & Training             & Validation           & Training            & Validation            \\ \hline
Nearest Neighbors                                 & 85                & 70                & 87                & 72                & 86                 & 72                 & 97                & 95                  \\ \hline
Linear SVM                                        & 88                & 83               & 87                & 84                & 86                 & 84                 & 98                & 98                  \\ \hline
poly SVM                                          & 94                & 78               & 94                & 79                & 93                 & 78                 & 99                & 97                  \\ \hline
sigmoid SVM                                       & 53                & 55               & 59                & 58                & 46                 & 48                 & 94                & 94                  \\ \hline
RBF SVM                                           & 99                & 80               & 99                & 82                & 99                 & 82                 & 100                   & 97                  \\ \hline
Gaussian Process                                  & 95                & 86               & 96                & 88                & 96                 & 88                 & 99                & 98                  \\ \hline
Decision Tree                                     & 50                 & 45               & 54                & 46                & 54                 & 52                 & 90                 & 90                   \\ \hline
Random Forest                                     & 78                & 75               & 77                & 75                & 75                 & 73                 & 97                & 96                  \\ \hline
Neural Net                                        & 100                   & 80                & 100                   & 80                 & 100                    & 80                  & 100                   & 97                  \\ \hline
AdaBoost                                          & 54                & 52               & 54                & 56                & 53                 & 55                 & 93                & 93                  \\ \hline
Gaussian Naive Bayes                              & 30                 & 31               & 56                & 48                & 41                 & 41                 & 83                & 84                  \\ \hline
Bernouill Naive Bayes                             & 29                & 27               & 37                & 31                & 37                 & 35                 & 84                & 85                  \\ \hline
QDA                                               & 78                & 65               & 84                & 73                & 83                 & 73                 & 97                & 96                  \\ \hline
Gradient Boosting                         & 99                & 76               & 100                   & 82                & 100                   & 82                 & 100                   & 97                  \\ \hline
\textbf{Voting Classifier}                   & \textbf{99}       & \textbf{87}      & \textbf{99}       & \textbf{88}       & \textbf{99}        & \textbf{88}        & \textbf{100}          & \textbf{98}         \\ \hline
\end{tabular}
\end{table*}

\begin{table*}[!h]
\caption{\textbf{Predictive Power: Tumor Types, Referential Signature~[28]} Training-Test  tumor types classification performance using the referential signature ($78$ genes) after retraining on entire Training-Validation set}
\centering
\begin{tabular}{|l|c|c|c|c|c|c|c|c|}
\hline
\multirow{2}{*}{Classifier}        & \multicolumn{2}{l|}{Balanced   Accuracy (\%)} & \multicolumn{2}{l|}{Weighted   Precision (\%)} & \multicolumn{2}{l|}{Weighted   Sensitivity (\%)} & \multicolumn{2}{l|}{Weighted   Specificity (\%)} \\ \cline{2-9} 
                                   & Training            & Test               & Training            & Test                & Training             & Test                 & Training            & Test                  \\ \hline
Linear SVM                         & 88                & 82               & 87                & 81                & 86                 & 81                 & 98                & 98                  \\ \hline
RBF SVM                            & 99                & 80               & 99                & 81                & 99                 & 81                 & 100                   & 97                  \\ \hline
Gaussian Process                   & 95                & 83               & 95                & 83                & 95                 & 83                 & 99                & 97                  \\ \hline
\textbf{Voting Classifier}    & \textbf{100}       & \textbf{85}      & \textbf{100}       & \textbf{89}       & \textb{100}        & \textbf{89}        & \textbf{100}          & \textbf{98}         \\ \hline
\end{tabular}
\end{table*}

Tables~VII and~VIII we present the performances for the random signature. Using the random signature only two classifiers were selected, fulfilling the used criteria. This signature reports the lowest performance compared to the other two signatures. 

\begin{table*}[!t]
\caption{\textbf{Predictive Power: Tumor Types, Random Signatures} Training-Validation  tumor types classification average performance over $10$ random signatures ($27$ genes each). Voting Classifier is composed of classifiers having reached a balanced accuracy above $80\%$ on validation.}
\centering
\begin{tabular}{|l|c|c|c|c|c|c|c|c|}
\hline
\multirow{2}{*}{Classifier} & \multicolumn{2}{l|}{Balanced   Accuracy (\%)}    & \multicolumn{2}{l|}{Weighted   Precision (\%)}   & \multicolumn{2}{l|}{Weighted   Sensitivity (\%)} & \multicolumn{2}{l|}{Weighted   Specificity (\%)} \\ \cline{2-9} 
                            & Training             & Validation           & Training             & Validation           & Training             & Validation           & Training             & Validation           \\ \hline
Nearest Neighbors           & 83+/-2          & 71+/-3          & 84+/-1          & 72+/-2          & 83+/-2          & 72+/-3          & 97+/-1          & 95+/-1          \\ \hline
Linear SVM                  & 8+/-1           & 78+/-2          & 78+/-1          & 77+/-2          & 77+/-2          & 75+/-2          & 97+/-0           & 97+/-0           \\ \hline
poly SVM                    & 94+/-1          & 79+/-2          & 93+/-2          & 78+/-2          & 92+/-2          & 78+/-2          & 99+/-0           & 97+/-0           \\ \hline
sigmoid SVM                 & 39+/-6          & 37+/-7          & 55+/-4          & 56+/-5          & 34+/-6          & 34+/-6          & 94+/-1          & 94+/-1          \\ \hline
RBF SVM                     & 9+/-1           & 8+/-2           & 88+/-2          & 8+/-2           & 88+/-2          & 8+/-2           & 98+/-0           & 97+/-0           \\ \hline
Gaussian Process            & 89+/-1          & 81+/-2          & 89+/-1          & 81+/-2          & 89+/-1          & 81+/-2          & 98+/-0           & 97+/-0           \\ \hline
Decision Tree               & 49+/-4          & 48+/-5          & 55+/-14          & 41+/-11          & 41+/-9          & 4+/-9           & 92+/-2          & 92+/-2          \\ \hline
Random Forest               & 76+/-2          & 75+/-3          & 75+/-2          & 73+/-3          & 73+/-2          & 71+/-3          & 97+/-0           & 96+/- 0          \\ \hline
Neural Net                  & 99+/-1          & 76+/-2          & 99+/-1          & 76+/-2          & 99+/-1          & 76+/-2          & 100+/-0            & 96+/-0           \\ \hline
AdaBoost                    & 56+/-5          & 53+/-5          & 57+/-4          & 55+/-5          & 53+/-6          & 51+/-7          & 93+/-1          & 93+/-1          \\ \hline
Gaussian Naive Bayes        & 26+/-6          & 28+/-7          & 57+/-6          & 55+/-8          & 41+/-5          & 42+/-6          & 81+/-3          & 81+/-4          \\ \hline
Bernouill Naive Bayes       & 28+/-7          & 28+/-7          & 42+/-4          & 34+/-8          & 38+/-5          & 38+/-5          & 86+/-3          & 86+/-3          \\ \hline
QDA                         & 60+/-11           & 57+/-10           & 69+/-6          & 65+/-6          & 50+/-15           & 48+/-14          & 95+/-1          & 95+/-1          \\ \hline
Gradient Boosting           & 100+/-0            & 78+/-3          & 100+/-0            & 80+/-2           & 100+/-0            & 80+/-2           & 100+/-0            & 97+/-0           \\ \hline
\textbf{Voting Classifier}  & \textbf{94+/-1} & \textbf{83+/-2} & \textbf{93+/-1} & \textbf{82+/-2} & \textbf{93+/-1} & \textbf{82+/-2} & \textbf{99+/-0}  & \textbf{98+/-1} \\ \hline

\end{tabular}
\end{table*}

\begin{table*}[!h]
\caption{\textbf{Predictive Power: Tumor Types, Random Signatures} Training-Test  tumor types classification average performance over $10$ random signatures ($27$ genes each) after retraining entire Training-Validation set}
\centering
\begin{tabular}{|l|c|c|c|c|c|c|c|c|}
\hline
\multirow{2}{*}{Classifier} & \multicolumn{2}{l|}{Balanced Accuracy (\%)}      & \multicolumn{2}{l|}{Weighted   Precision (\%)}   & \multicolumn{2}{l|}{Weighted   Sensitivity (\%)} & \multicolumn{2}{l|}{Weighted   Specificity (\%)} \\ \cline{2-9} 
                            & Training             & Test                 & Training             & Test                 & Training             & Test                 & Training             & Test                 \\ \hline
RBF SVM                     & 90+/-1           & 79+/-2          & 88+/-2          & 79+/-1          & 88+/-2          & 78+/-2          & 98+/-0           & 97+/-0           \\ \hline
Gaussian Process            & 89+/-1          & 80+/-1           & 89+/-1          & 80+/-1           & 89+/-1          & 81+/-1          & 98+/-0           & 97+/-0           \\ \hline
\textbf{Voting Classifier}  & \textbf{96+/-5} & \textbf{84+/-2} & \textbf{95+/-5} & \textbf{87+/-3} & \textbf{94+/-7} & \textbf{86+/-4} & \textbf{99+/-1}  & \textbf{97+/-1}  \\ \hline
\end{tabular}
\end{table*}